\pdfoutput=1

\documentclass[graybox, nosecnum]{svmult}

\usepackage{mathptmx}       
\usepackage{helvet}         
\usepackage{courier}        
\usepackage{type1cm}        
\usepackage{makeidx}         
\usepackage{graphicx}        
\usepackage{multicol}        
\usepackage[bottom]{footmisc}
\usepackage{hyperref}        
\usepackage{soul}            
\hypersetup{colorlinks=true,urlcolor=blue}
\usepackage[square,numbers]{natbib}
\makeindex             

\begin{document}
\title*{XMM-Newton}
\titlerunning{XMM-Newton}
\author{Norbert Schartel\thanks{corresponding author (\email{Norbert.Schartel@esa.int})},
  Rosario Gonz{\'a}lez-Riestra, Peter Kretschmar, Marcus Kirsch, Pedro Rodr{\'i}guez-Pascual, Simon Rosen,
  Maria Santos-Lle{\'o}\thanks{corresponding author (\email{Maria.Santos-Lleo@esa.int})},
   Michael Smith, Martin Stuhlinger and Eva Verdugo-Rodrigo}
  \authorrunning{Schartel}
\institute{Norbert Schartel, Peter Kretschmar, Maria Santos-Lle{\'o}  and Eva Verdugo-Rodrigo \at European Space Agency (ESA), European Space Astronomy Centre (ESAC), Camino Bajo del Castillo s/n, 28692, Villanueva de la Ca{\~{n}}ada, Madrid, Spain 
  \and Rosario Gonz{\'a}lez-Riestra, Pedro Rodr{\'i}guez-Pascual, Simon Rosen and Martin Stuhlinger \at Serco Gesti{\'o}n de Negocios S.L., ESAC, Camino Bajo del Castillo s/n, 28692, Villanueva de la Ca{\~{n}}ada, Madrid, Spain  
\and Marcus Kirsch \at ESA, European Space Operations Centre (ESOC),  Robert-Bosch-Str. 5, 64293 Darmstadt, Germany 
\and Michael Smith \at Telespazio, ESAC, Camino Bajo del Castillo s/n, 28692, Villanueva de la Ca{\~{n}}ada, Madrid, Spain 
}

\maketitle
\abstract{
The X-ray Multi-mirror Mission (XMM-Newton) provides simultaneous non-dispersive spectroscopic
X-ray imaging and timing, medium resolution dispersive X-ray spectroscopy and optical/UV imaging,
spectroscopy and timing. \\
In combination, the imaging cameras offer an effective area over the energy range
from 150 eV to 12 keV of up to 2500 cm$^2$ at 1.5 keV and $\sim$1800 cm$^2$ at 5 keV.
The gratings cover an energy range from 0.4 keV to 2.2 keV with a combined effective area
of up to 120 cm$^2$ at 0.8 keV. \\
XMM-Newton offers unique opportunities for a wide variety
of sensitive X-ray observations accompanied by simultaneous optical/UV measurements.
The majority of XMM-Newton's observing time is made available to the astronomical
community by peer-reviewed Announcements of Opportunity. \\
The scientific exploitation of XMM-Newton data is aided by an observatory-class X-ray facility which provides analysis software, pipeline processing, calibration and catalogue generation.
Around 380 refereed papers based on XMM-Newton data are published each year with a high fraction of papers reporting transformative scientific results. 
}
\section{Keywords} 
XMM-Newton, Reflection Grating Spectrometers, European Photon Imaging Camera, Optical Monitor, X-ray detectors, X-ray telescopes, joint programmes, Target of Opportunity.

\section{Introduction}
\label{sec:Intr}

The design and construction of XMM-Newton (X-ray Multi-Mirror Mission Newton, \cite{Jansen2001})  and its instruments utilized
the extensive experience in observing X-ray astrophysical sources that European and United States (US) scientists accumulated during the previous few years. 

The European X-ray Observatory SATellite (EXOSAT) which was operational between
1983 and 1986, was  the European Space Agency's (ESA) direct predecessor to XMM-Newton.
Two further large European X-ray satellite missions preceded XMM-Newton:
the German-British-US ROSAT (R{\"o}ntgensatellit)
(\cite{Truemper1983}), operational from 1990 to 1999, and the
Italian-Dutch BeppoSAX (Satellite per Astronomia a raggi X,
(\cite{Boella1997})), which was operational between 1996 and 2002.

XMM-Newton was proposed to ESA as the X-ray Multi-Mirror (XMM) astronomy mission
by
J.A.M. Bleeker\footnote{Leiden, The Netherlands},
A.C. Brinkman\footnote{Utrecht, The Netherlands},
J.L. Culhane\footnote{Mullard Space Science Laboratory, United Kingdom},
L. Koch\footnote{Saclay, France},
K.A. Pounds\footnote{Leicester,  United Kingdom},
H.W. Schnopper\footnote{Lyngby, Denmark},
G. Spada\footnote{Bologna, Italy},
B.G. Taylor\footnote{ESA, The Netherlands}
and
J. Tr{\"u}mper\footnote{Garching, Germany}
in November 1982. 
An important milestone in the formulation of the scientific and technical
requirements for XMM-Newton was the ESA workshop in Lyngby, Denmark in June 1985
(\cite{Lyngby1985}), where in addition to the X-ray capabilities the importance of simultaneous
optical/UV coverage was emphasized.
Comprehensive technical development in European and
US industries and institutes was necessary to fulfil the 
scientific requirements for XMM-Newton. Significant
achievements took place, i.e., within X-ray optics development with mirror replication from
super-polished gold-coated mandrels using nickel electroforming, X-ray CCD technology 
with the pn CCDs specifically developed for XMM-Newton and 
the development of X-ray reflection gratings technology.

\begin{figure}[ht]
\centering
\includegraphics[scale=2.8]{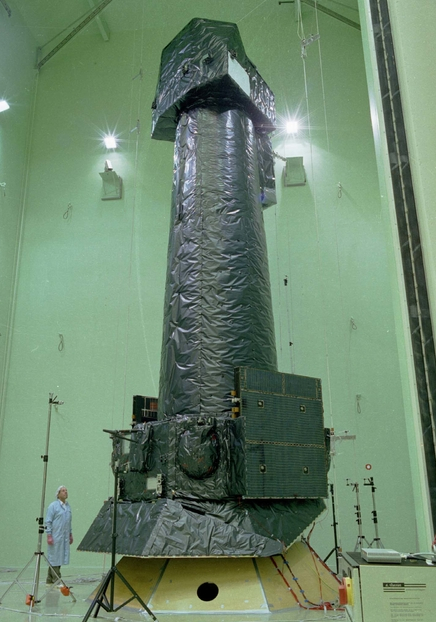}
\caption{
The XMM-Newton Flight Model spacecraft after completion of acoustic testing inside the Large European Acoustic Facility  at the European Space Research and Technology Centre (ESTEC), Noordwijk, the Netherlands}
\label{fig:spacecraft2}       
\end{figure}
\begin{figure}[h]
\centering
\includegraphics[width=120mm]{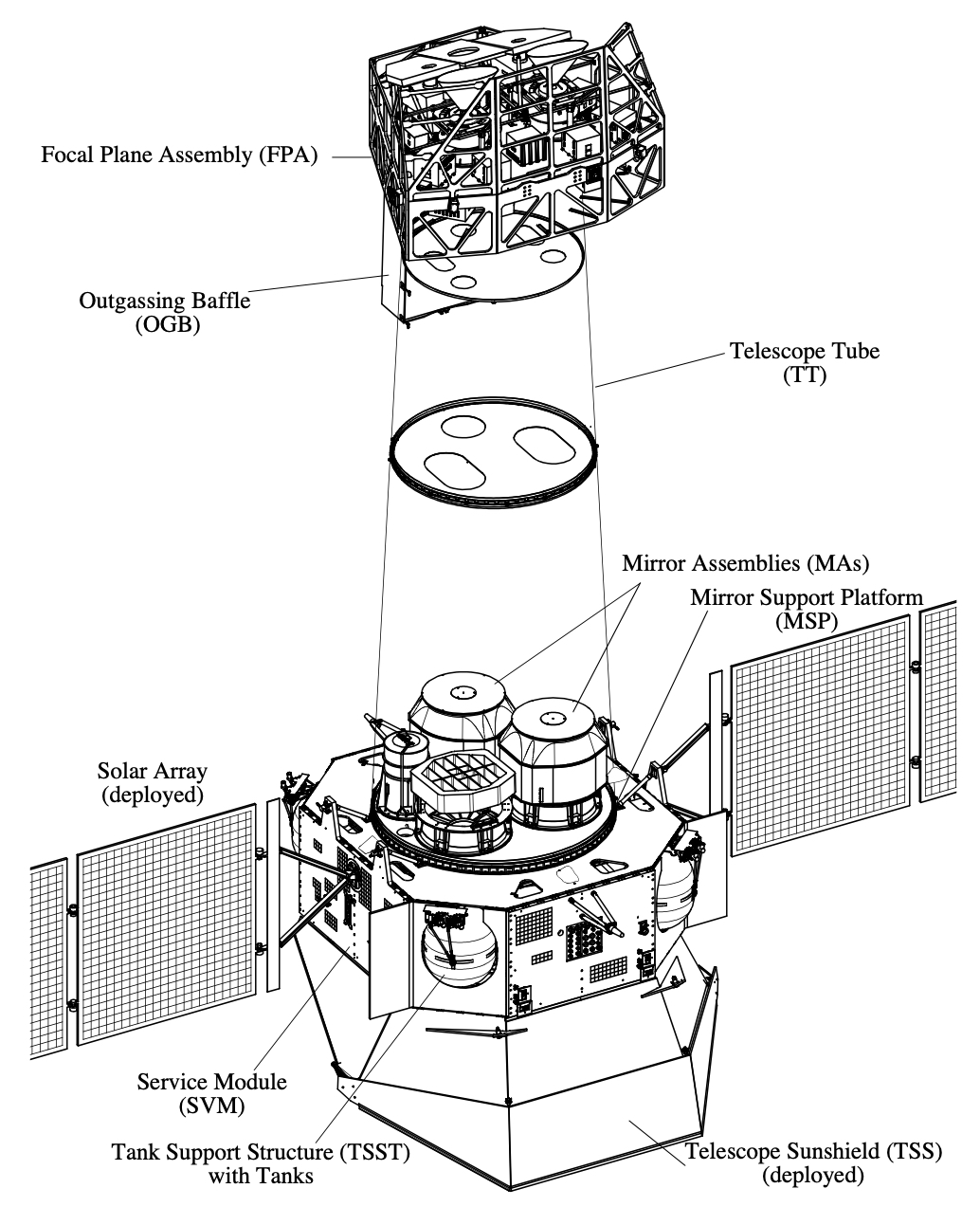}
\caption{A technical drawing of the XMM-Newton spacecraft from the Original Dornier Users Manual from 1999}
\label{fig:spacecraft}       
\end{figure}

XMM-Newton carries three X-ray telescopes. Located behind two of the telescopes are two Reflection Grating Spectrometers (RGS, (\cite{denHerder2001})) which intercept about half of the incident photons. 
The remaining photons are focused onto two MOS cameras (\cite{Turner2001}).
The third telescope focuses all light onto the pn camera (\cite{Strueder2001}).
The two MOS cameras and the pn camera constitute the European Photon Imaging Camera (EPIC).
The sixth instrument onboard XMM-Newton is the Optical Monitor (\cite{Mason2001}).
All six instruments normally observe simultaneously.

XMM-Newton was the second cornerstone of the ESA Horizon 2000
program and was launched by an Ariane V rocket on December 10, 1999.
It was ESA's first large observatory-class mission, where the scientific support (peer review proposal selection, calibration, analysis software and archive) was planned for and provided from the start.

In the following we give an overview of the spacecraft (page~\pageref{sec:Craft}), the X-ray telescopes (page~\pageref{sec:Telescopes}), and the instruments (EPIC  (page~\pageref{sec:EPIC}), RGS (page~\pageref{sec:RGS}), and OM (page~\pageref{sec:OM})) as well as how the ground segment is organized  (page~\pageref{sec:ground}), how the observing program is implemented (page~\pageref{sec:obs}) and how the science data can be analyzed
(page~\pageref{sec:data}),  the observing strategy (page~\pageref{sec:stra}) and its evolution through the years and the scientific output as reflected, e.g., in published papers.


\section{\textit{The Spacecraft}}
\label{sec:Craft}

The three-axis stabilized, 3.8 tonne spacecraft, see Fig.~\ref{fig:spacecraft2} and Fig.~\ref{fig:spacecraft},
has a pointing accuracy of one arcsec. It consists of three main sections: a 7 m-long telescope tube between the squarish service module which houses the instruments and the telescope module accommodating the three telescopes. This gives the spacecraft a total length of 10.8 m. Its pair of solar panels have a 16 m span.

The spacecraft was built by Dornier (now Airbus Defence and Space) from 1994 to 1999.
Its main subsystems are the Attitude and Orbit Control System including the Reaction Control System  (AOCS and RCS), On Board Data Handling System (OBDH), the Power and Thermal Systems (EPS, TCS), and the Radio Frequency System (RF). All subsystems have their own redundancy concepts.

The AOCS provides three-axis stabilization during all modes. The AOCS architecture is formed around the Attitude Control Computer (ACC), running the software for mode control and the attitude and thrust control laws. The AOCS uses a Star Tracker and Fine Sun Sensor to provide the absolute reference. The Star Tracker is a small telescope with 3$^{\rm o}$ $\times$ 4$^{\rm o}$ field of view (FOV) and a thermo-electrically cooled CCD detector. The Fine Sun Sensors deliver pitch and roll information and their FOV is
$\pm$45$^{\rm o}$ per sensor. Backup sensors such as Inertial Measurement Units are also used for control in case of main sensors unavailability such as the Fine Sun Sensor during the eclipse and the Star Tracker when its FOV is blinded or the guide star is lost. Reaction wheels are the primary actuators for attitude control. In the classic three-wheel mode operations scenario, any three out of four reaction wheels are used for active control, each one with a net torque of 0.2 Nm and 40 Nms momentum capacity. The reaction wheel that is not used for active control was originally switched off and only used in cold redundancy. The failure detection and correction is performed using hardware only.

The RCS is a monopropellant propulsion system utilizing hydrazine (N$_2$H$_4$) in blow-down mode with helium as the pressurant. The propellant is stored in four surface tension tanks placed around the central cone element of the service module structure of the spacecraft. Eight monopropellant 20 Newton thrusters, divided into two redundant branches, each consisting of four thrusters, are used for delta-V maneuvers, attitude control, and reaction wheel desaturation.

\begin{figure}[h]
\centering
\includegraphics[width=70mm]{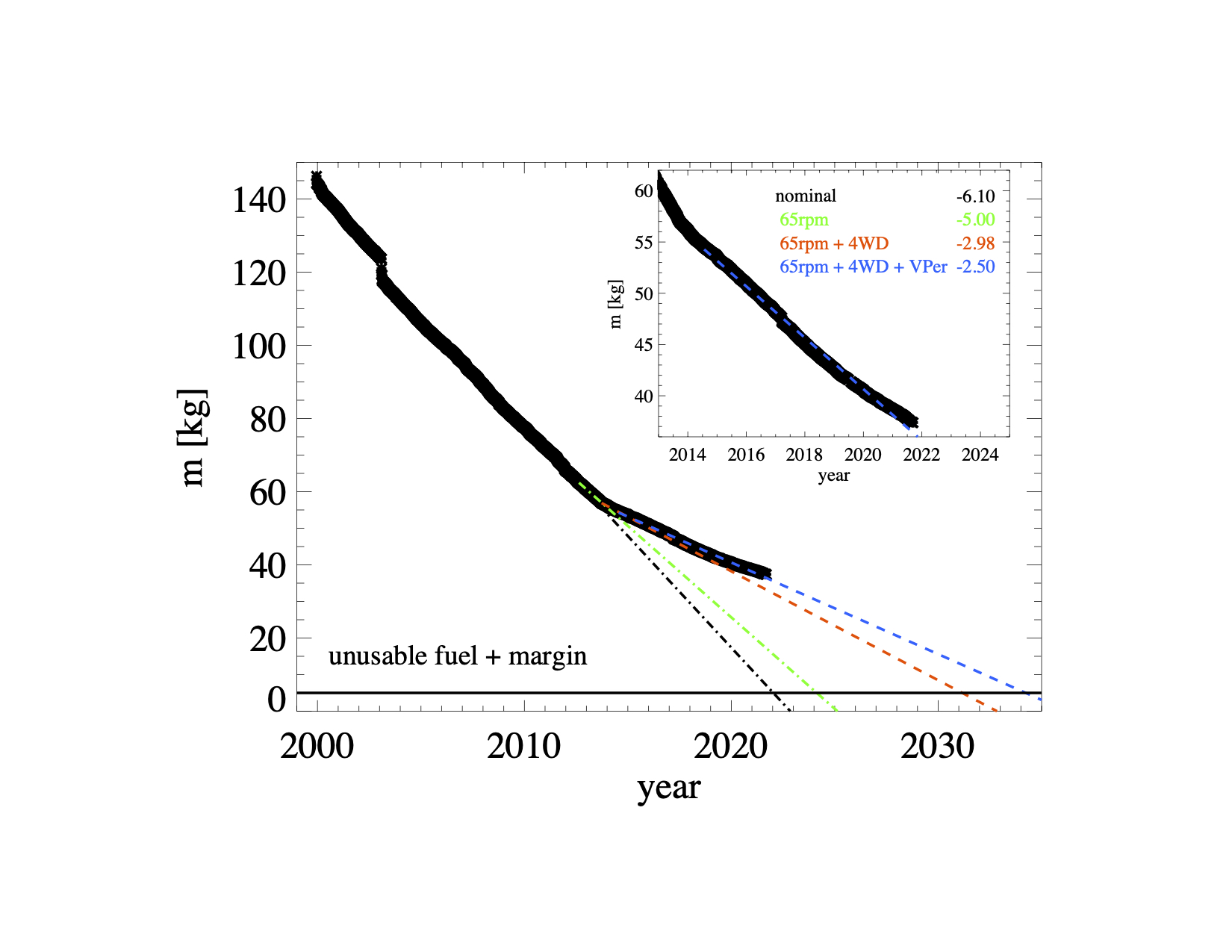}
\caption{XMM-Newton fuel consumption calculated via the flight dynamics book-keeping method. The extrapolation does not include any contingency cases. As of 2013 the annual fuel consumption was reduced by more than a factor of two}
\label{fig:fuel_consumption}       
\end{figure}

The functional objective of the OBDH subsystem is to provide command and telemetry capability allowing the overall spacecraft control and to provide clock and timing facilities for the other subsystems. 
It consists of a CDMU, two Remote Terminal Units (RTUs) and six Digital Bus Units (DBUs). The RTUs connect via standard discrete interfaces to all users without intelligence (i.e. microprocessors), e.g., the EPS, RF, and TCS. The RTUs handle all instrument functions and the AOCS where no microprocessor involvement of the relevant unit is available.
The OBDH system is based on the implementation of the ESA Packet Telecommand Standard PSS-04-107, ESA Packet Telemetry Standard PSS-04-106 and an ESA OBDH Data Bus as main path for telecommand distribution and telemetry acquisition. The users are distinguished between Packet Terminals and Non-Packet Terminals.

The EPS provides electrical power to all subsystems. During sunlight power is provided by the fixed, two-wing deployable solar array and in eclipse by two nickel-cadmium batteries. Both energy sources are conditioned to provide a single 28 V DC regulated bus. The solar array generates a maximum power of 1.8 kW.

The TCS features a combination of active and passive thermal control systems originally without any onboard control software. This means that all the heater lines are either online controlled from ground or thermostat-controlled. The ground commands open or close transistor switches installed within the service module power distribution units, focal plane assembly power distribution units and mirror thermal control unit in the power lines between heater and latching current limiters. During ground contact outages, the heater lines are controlled by bi-metal mechanical thermostats in the satellite. 

The Radio Frequency (RF) subsystem is composed of three main blocks which are two Low Gain Antennas, two transponders, and a Radio Frequency Distribution Network with two switches to connect the transponders to the antennae. An S-Band transponder comprises three main modules: a diplexer, a receiver, and a transmitter. The receiver assures the reception of signal in the range of 2025 to 2120 MHz (S-band) and the phase demodulation of the telecommand signal and the ranging tones. The transmitter performs the modulation of the telemetry video signal and the ranging tones, as well as the power amplification of the output signal. The diplexer allows simultaneous operation of the receiver and the transmitter with just a single RF connection. This configuration allows an effective data rate of 80 kb s$^{-1}$ for the downlink and 2 kb s$^{-1}$ for the uplink.

XMM-Newton has no onboard data storage capacity, so all data are immediately down-linked to the ground in real time. The mission requires control 24/7, 365 days per year through a live ground station connection, since the spacecraft has only limited automation.

After a loss of contact to the spacecraft for 5 days in 2008, due to one unreliable RF subsystem antenna switch, the mission was recovered and the mission operations concept revised to use both the prime and redundant transponders for communication. This method has not reduced the science return \cite{Kirsch2010}.

In 2013, the AOCS was re-programmed in order to also use the backup reaction wheel in a new operations scenario called "4 wheel drive" (4WD). Since 2013 the degree of freedom that was introduced by this change allows the fuel consumption to be reduced by a factor of two, see Fig.~\ref{fig:fuel_consumption}, and provides as well mitigation measures against a degradation effect on two of the reaction wheels \cite{Pantaleoni2012} \cite{Kirsch2014}. Technically, the observatory currently has sufficient resources to operate until 2031

The onboard software has also been updated several times to improve and streamline mission operations and to support the provision of fuel in the near depletion regime, allowing now fine thermal control of some subsystems as well as switch state protection of some Heater Power Lines \cite{Kirsch2014}.

\begin{figure}[t]
  \begin{minipage}[t]{5.5cm}
    	\begin{center}
	\includegraphics[angle=0,scale=0.23]{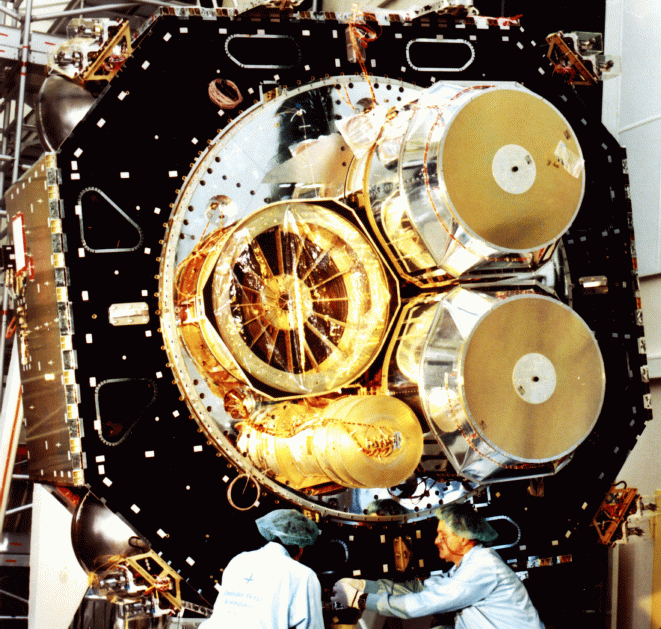}
	\caption{\label{Fig:mirr2}
          The XMM-Newton mirror modules on the backside of the XMM-Newton service module.
          One spider is visible on the left side, while the two other modules are covered}	 
	\end{center}      
    \end{minipage}
  \hfill
  \begin{minipage}[t]{5.5cm}
    \begin{center}
		\includegraphics[angle=0,scale=0.245]{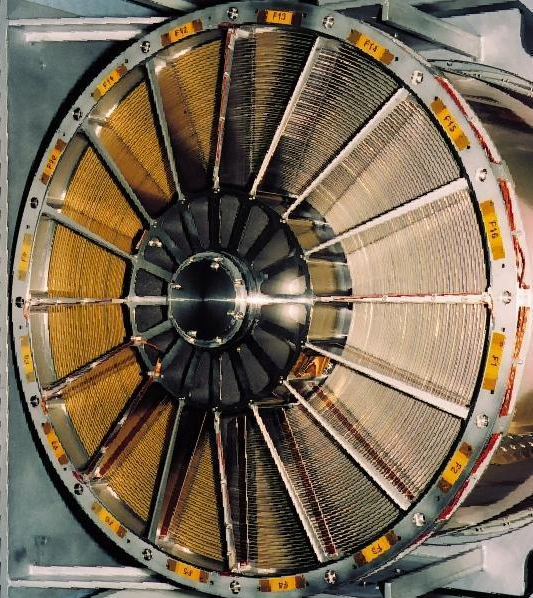}
		\caption{\label{Fig:mirr1}
                  One spider of the  XMM-Newton mirror module carries a full set of 58 flight mirror shells}
	\end{center}
    \end{minipage}
\end{figure}


\begin{figure}[thbp]
\begin{center}
\includegraphics[scale=0.45]{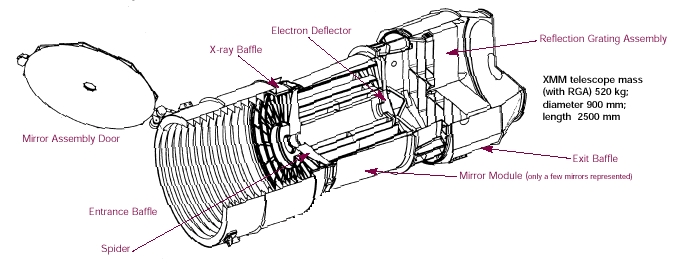}
  \caption{X-ray telescope configuration}
\label{fig:mirrors}
\end{center}
\end{figure}

\section{\textit{X-Ray Mirrors}}
 \label{sec:Telescopes}

Each of the three X-ray telescopes on board XMM-Newton consists of 58 Wolter I grazing-incidence mirrors which are nested in a coaxial and cofocal configuration,
see; \cite{Gondoin1998a},  \cite{Gondoin1998b}, \cite{Aschenbach2001}, \cite{Jansen2001} and \cite{Ebrero2021},
see Fig.~\ref{Fig:mirr2}, Fig.~\ref{Fig:mirr1}, and Fig.~\ref{fig:mirrors}.
The design of the optics was driven by the requirement of obtaining the highest possible effective area over the 0.2-12.0 keV energy range, with particular emphasis in the region around 7 keV. Thus, the mirror system had to utilize a very shallow grazing angle of 30' in order to provide sufficient reflectivity at high energies. The telescopes' focal length is 7.5 m, and the diameter of the largest mirrors is 70 cm, to be compatible with the shroud of the launcher. 

The mirrors were replicated from super-polished gold-coated mandrels using a nickel electroforming technique. The 58 mirrors of each telescope are bonded onto their entrance aperture to the 16 spokes of a single spider. An electron deflector is located in the exit aperture which produces a circumferential magnetic field to prevent low-energy electrons reflected by the mirrors reaching the focal plane detectors. X-ray baffles, consisting of 2 sieve plates each with 58 annular apertures, are located in front of the mirror systems. They act as collimators and considerably reduce the amount of straylight in the FOV of the focal plane cameras.

The spider is connected to the support platform via an aluminum interface structure (the MIS: Mirror Interface Structure) consisting of an outer cylinder and an interface ring. On two of the modules, the ring interfaces the mirror module to a Reflection Grating Assembly (RGA). To minimize the mechanical deformation of the mirrors and therefore the optical degradation, the flatness of the interface between the spider and the MIS is less than 5 micron. 

The first critical parameter determining the quality of an X-ray mirror module is its ability to focus photons. This is one of XMM-Newton's major strengths: the core of its on-axis point-spread function (PSF) is narrow and varies little over a wide energy range (0.1-6 keV). Above 6 keV, the PSF becomes only slightly more energy dependent.

\begin{figure}[htbp]
  \begin{center}
    
\includegraphics[scale=0.40]{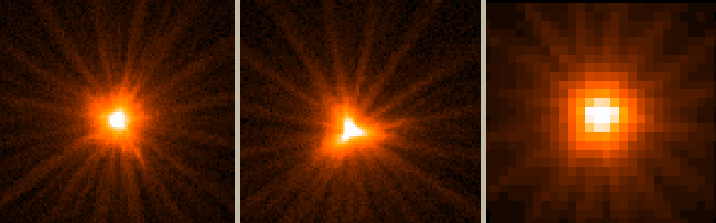}
\caption{On-axis point-spread function (PSF) of the EPIC-MOS1, EPIC-MOS2 and EPIC-pn X-ray telescopes (left to right) registered on the same source with each EPIC-MOS camera in Small Window mode, and the EPIC-pn camera in Large Window mode}
\label{fig:onpsf}
\end{center}
\end{figure}
\begin{figure}[htbp]
\begin{center}
\includegraphics[scale=0.40]{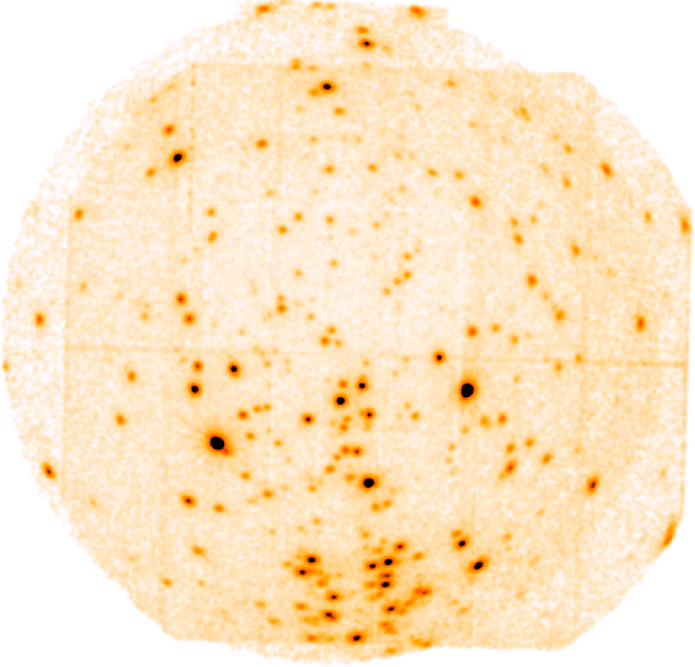}
\caption{The dependence of the X-ray PSF on position in the FOV. This image was made from an observation towards the Orion molecular cloud. EPIC-pn, EPIC-MOS1 and EPIC-MOS2 exposures have been merged and exposure corrected. The image has been slightly smoothed with a Gaussian of 10" FWHM. The intensity scale is square root}
\label{fig:offpsf}
\end{center}
\end{figure}

The details of the PSF shapes and energy dependencies differ between the respective telescopes. 
As an example, Fig.\ref{fig:onpsf} shows the in-orbit on-axis PSF of the EPIC-MOS1, EPIC-MOS2, and EPIC-pn X-ray telescopes, measured on the same source. This figure is primarily provided to show the shape of the PSF, with e.g., the radial substructures caused by the spiders holding the mirror shells.

The PSF of the X-ray telescopes depends on the source off-axis angle, i.e., its distance from the center of the FOV. It also depends slightly on the source azimuth within the FOV. In Fig.\ref{fig:offpsf} the dependence of the shape of the XMM-Newton X-ray PSF on the position within the FOV is presented. The PSF at large off-axis angles is elongated due to off-axis aberration (astigmatism).


\section{\textit{European Photon Imaging Camera (EPIC)}}
 \label{sec:EPIC}

The European Photon Imaging Camera (EPIC) offers the possibility of performing extremely sensitive imaging observations over the telescopes' circular FOV of 30' diameter in the 0.15 - 15keV energy range with moderate spectral ($E/\Delta E \sim$ 20--50) and angular resolution (6'' FWHM point-spread function). One camera can be operated with a time resolution as fast as 7 $\mu$s.

The EPIC consists of three imaging spectrometers each of which is located at the focus of one of the three coaligned X-ray telescopes. Two of the cameras consist of an array of seven EPIC-MOS-type
 Charge-Coupled Devices (CCDs) 
\cite{Turner2001} and are 
located behind the X-ray telescopes which share their optical paths with the RGS \cite{denHerder2001}. The grating 
assemblies intercept the light path such that about 44\% of the original incident flux reaches the EPIC-MOS cameras. 
The third EPIC camera system employs a monolithic pn-CCD array \cite{Strueder2001} and is located
at the focus of an unobstructed X-ray telescope. Each EPIC instrument is fitted with an identical filter wheel containing several
X-ray transmissive but optical light blocking filters as well as fully open and fully closed positions. The EPIC instrument also includes the 
Radiation Monitoring System, which detects the ambient electron and proton fluxes and is used as part of the warning system to protect the cameras in case of harmful radiation levels.

\begin{figure}
\centering
\includegraphics[width=1.0\linewidth]{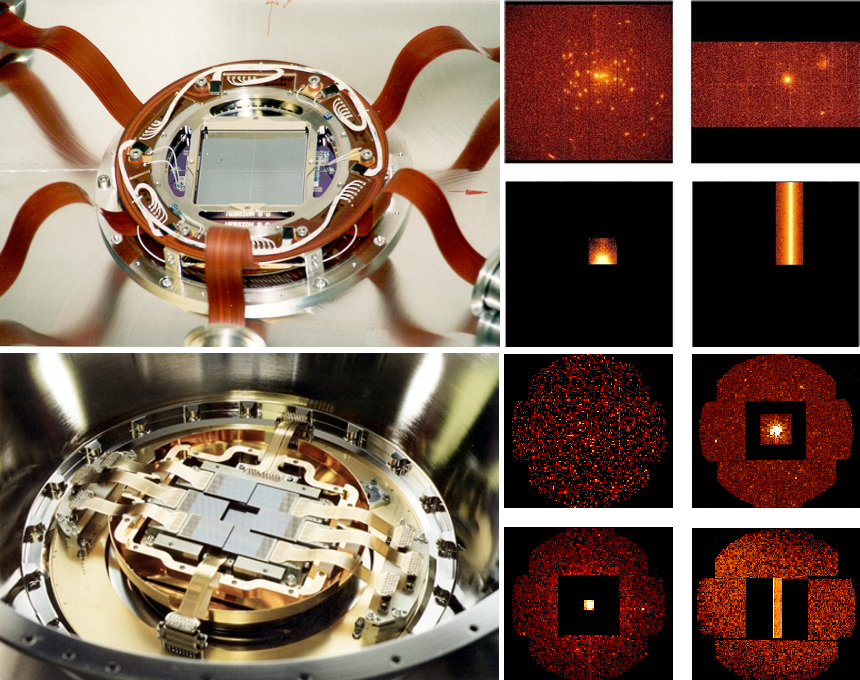}
\caption{Images of the CCD detectors EPIC-pn (top) and EPIC-MOS as well as examples of the FOVs using the different observation modes - Full Frame, Large Window, Small Window, and Timing
\label{fig:epicdetectors_modes}}
\end{figure}

\subsection{\textit{The Instrument}}
Each EPIC-MOS camera contains seven front-illuminated CCDs; see Fig.~\ref{fig:epicdetectors_modes}.
These are three-phase frame transfer devices of high
resistivity epitaxial silicon with an open-electrode structure. The central CCD is located at the telescope focus while the 
outer six CCDs are stepped towards the mirror to follow approximately the focal plane curvature to help with maintaining focus 
for off-axis sources. The CCD imaging area is 2.5 $\times$ 2.5 cm, and contains 600 $\times$ 600, 40 micron square pixels; one pixel 
covers 1.1 $\times$ 1.1'' on the FOV; 15 pixels cover the mirror PSF half energy width of 15''.
The mosaic of seven CCDs covers the focal plane of 62 mm, equivalent to 28.4 arcmin, in diameter. 
The two EPIC-MOS cameras are arranged on the spacecraft in a mutually orthogonal 
layout, such that the gaps in sky coverage between the outer CCDs of one unit are covered by the other.

The EPIC-pn detector was specifically designed for the XMM-Newton X-ray telescope performance in terms of angular resolution, FOV, 
and energy pass band. The EPIC-pn camera array consists of 12 individually operated 3 $\times$ 1 cm back-illuminated pn-CCDs, 
monolithically implanted on a single wafer; see Fig.~\ref{fig:epicdetectors_modes}.
The instrument is subdivided into four individual quadrants of three pn-CCD subunits 
with a format of 200 $\times$ 64 pixels operated in parallel. 
The resulting imaging area of 6 $\times$ 6 cm covers approximately 97\% of the telescope FOV; about 6 cm$^2$ 
of the CCD sensitive area lies outside the FOV and allows instrumental background studies. The CCD pixel size
is 150 $\times$ 150 microns (4.1'') with a position resolution of 120 microns, resulting in an equivalent angular resolving
capability for a single photon of 3.3''. The focal point of the X-ray telescope is located on CCD~0 of quadrant 1.

Since launch, EPIC has suffered five events, ascribed to micrometeoroid impacts along the boresight, which have resulted in permanent damage to the detectors. 
In three cases the lasting effect was limited to the appearance of individual bright pixels. 
However, two events resulted in more extensive damage: the loss of EPIC-MOS1 CCD~6 and the appearance of a hot column 
passing very close to the EPIC-MOS1 boresight, and the loss of EPIC-MOS1 CCD~3.

The EPIC cameras allow several data acquisition modes to accommodate a range of source fluxes and to allow for fast-timing 
measurements, see Fig.~\ref{fig:epicdetectors_modes}.
With respect to the standard Full Frame imaging mode, the CCD readout speed can be increased at the cost of a smaller area of the
window to be read out. Timing mode (and also Burst mode for EPIC-pn) allow the fastest readout cycles by sacrificing the spatial
information in the readout direction.
In the case of EPIC-MOS, the various modes affect the central CCD only while the six peripheral CCDs remain
in standard Full Frame imaging mode. 
The EPIC-pn camera CCDs can be operated in common modes across all quadrants for Full Frame, Extended Full Frame and 
Large Window modes, or just with the single focal point CCD being read out for Small Window, Timing and Burst modes.
The duty cycle efficiency of most modes is 95--100\%. However, for the EPIC-pn Small Window mode and, especially, Burst mode the 
efficiencies are lower, at 71.0\% and 3.0\%, respectively. A summary of the basic characteristics of the science modes is given in 
Table~\ref{Table:EPICScienceModes}.

\begin{center}
  \begin{table}
   \begin{center} 
\begin{tabular}{| l | l | l |}\hline
EPIC-MOS (central CCD; pixels)        & Time       & Max. point source count rate (flux) \\
{[}1 pixel = 1.1"]                    & resolution & [s$^{-1}$] ([mCrab])                \\\hline \hline
Full frame (600 $\times$ 600)	      & 2.6 s      & 0.50 (0.17)                         \\
Large window (300 $\times$ 300)        & 0.9 s      & 1.5 (0.49)                          \\
Small window (100 $\times$ 100)        & 0.3 s      & 4.5 (1.53)                          \\
Timing uncompressed (100 $\times$ 600) & 1.75 ms    & 100 (35)                            \\\hline 
\end{tabular}
\vspace{0.5cm}
\begin{tabular}{| l | l | l |}\hline
EPIC-pn (array or 1 CCD; pixels)      & Time       & Max. point source count rate (flux) \\
{[}1 pixel = 4.1"]                    & resolution & [s$^{-1}$] ([mCrab])                \\\hline \hline
Full frame (376 $\times$ 384)          & 73.4 ms    & 2 (0.23)                            \\
Extended full frame (376 $\times$ 384) & 199.1 ms   & 0.7 (0.09)                          \\
Large window (198 $\times$ 384)        & 47.7 ms    & 3 (0.35)                            \\
Small window (63 $\times$ 64)          & 5.7 ms     & 25 (3.25)                           \\
Timing (64 $\times$ 200)               & 0.03 ms    & 800 (85)                            \\
Burst (64 $\times$ 180)                & 7 $\mu$s   & 60,000 (6300)                        \\\hline
\end{tabular}
\caption{Summary of the basic characteristics of the EPIC science modes.}
\label{Table:EPICScienceModes}
\end{center}
   \end{table}
\end{center}

\begin{figure}
\centering
\includegraphics[width=0.95\linewidth]{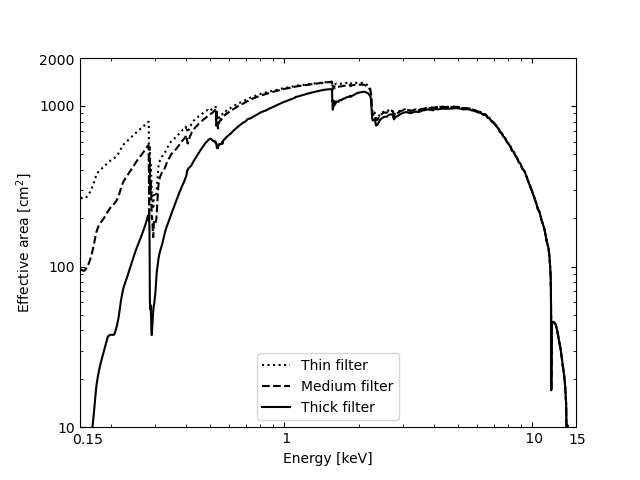}
\includegraphics[width=0.95\linewidth]{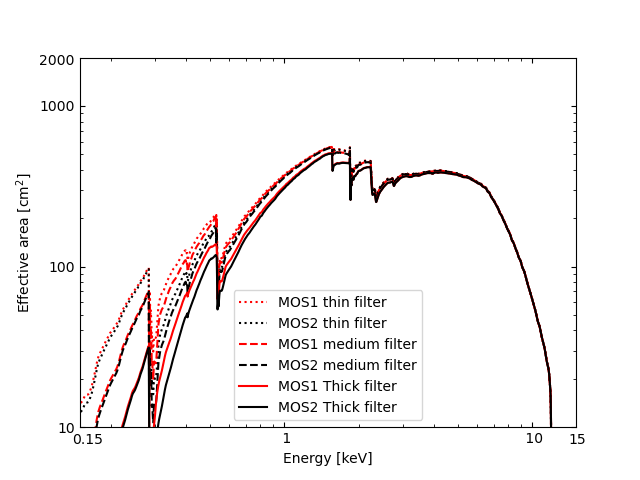}
\caption{Current total effective areas of the EPIC-pn (top) and EPIC-MOS (bottom). Due to contamination, probably by hydrocarbons, the EPIC-MOS2 effective area has decreased slightly at lower energies over the mission duration compared to the EPIC-MOS1
\label{fig:epic-effective-areas}}
\end{figure}

As the EPIC detectors are not only sensitive to X-ray photons but also to visible and UV light, the cameras include aluminized optical blocking filters. 
Their design is a compromise between the need to prevent optical and UV photons from reaching the CCD plane, and the need to absorb as few X-ray photons as possible, especially at the lowest X-ray energies.
There are four filters in each EPIC camera.
Two are thin filters made of 1600 {\AA} of unsupported polyimide film with a single 400 {\AA} layer of aluminum evaporated on one side.
A medium filter is of similar construction but with an 800 {\AA} layer of deposited aluminum.
The thick filter is constructed on an unsupported 3300 {\AA} thick polypropylene film with 1100 {\AA} of aluminum and 450 {\AA} of tin 
evaporated on the film to block the UV that would otherwise be transmitted
by the polypropylene. The filters are self-supporting and 76 mm in diameter. 
Two additional positions on the filter wheel are occupied by the closed (1.05 mm of aluminum) and open positions, respectively.
The former is used to protect the CCDs from soft protons, while the open position could in principle be used for observations where the light flux is very low and no filter is needed; see Fig. \ref{fig:epic-effective-areas}.

\subsection{\textit{Scientific Performance}}

The overall EPIC-MOS quantum efficiency (QE) is smooth function with prominent features at the Si and O edges. 
The low-energy QE varies somewhat from CCD to CCD.
It has become apparent that over the course of the mission the response of the EPIC-MOS cameras (primarily EPIC-MOS2) is deteriorating 
below $\sim$1 keV. It is suspected that this is due to a steady buildup of contaminant which has adhered to the surface of the 
cameras, absorbing a fraction of the incoming photons. This development is being closely monitored and the time-dependent change in 
efficiency is included in the instrument calibration.
The low-energy redistribution function (RMF) of the EPIC-MOS CCDs has a complex shape consisting of a main photo peak combined with
a low-energy shoulder, the latter becoming dominant towards the very lowest energies. In-flight measurements show it to be both 
temporally and spatially dependent, with the most pronounced effects around the bore-sight location.
The energy resolution of the EPIC-MOS cameras showed a significant temporal degradation in the early stage of the mission.
However, in November 2002 the cameras were cooled to run at a lower operating temperature,
thus significantly restoring the energy resolution and reducing its subsequent rate of 
degradation.

The EPIC-pn QE low-energy response is limited by the radiation entrance window and properties of the Si L-edge, whereas the 
high-energy response is determined by the fully depleted 300 $\mu$m depth of silicon. Further prominent features are due to 
absorption in the SiO$_2$ passivation layer (at 528 eV) and the X-ray absorption fine structure around the Si K-edge (at 
1.838 keV). The absolute QE was calibrated on ground at synchroton facilities under conditions comparable to space operations,
and further in-flight calibration measurements have shown it to be very stable over the course of the mission.
In-flight measurements of the energy resolution show a steady degradation by about 2.5 eV/year at 6 keV for Full Frame mode data at the boresight location.

\begin{figure}[t]
  \begin{minipage}[t]{5.5cm}
    	\begin{center}
	\includegraphics[angle=0,scale=0.2]{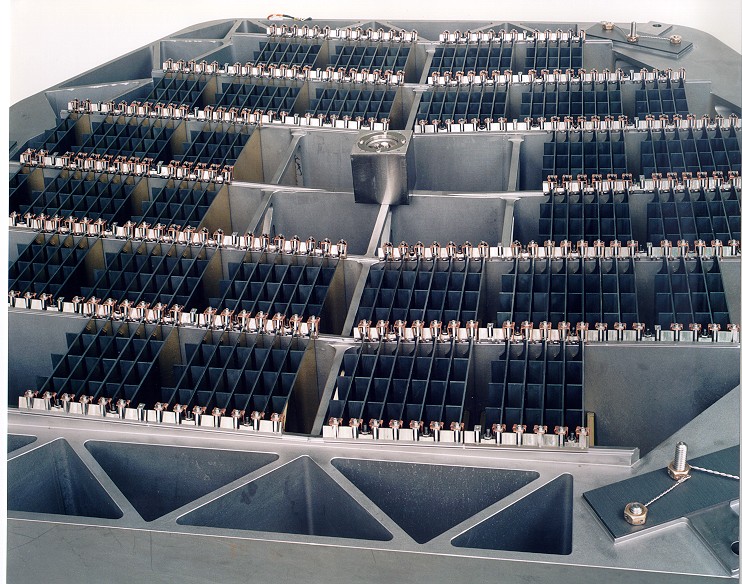}
	\caption{\label{Fig:RGA}
          View of one of the Reflection Grating Arrays}	 
	\end{center}      
    \end{minipage}
  \hfill
  \begin{minipage}[t]{5.5cm}
    \begin{center}
		\includegraphics[angle=0,scale=0.35]{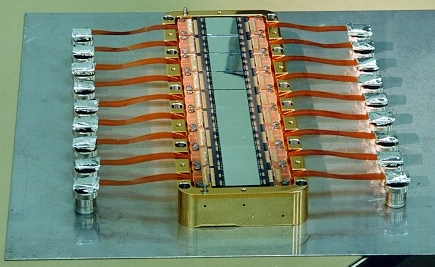}
		\caption{\label{Fig:RFC}
                  View of one of the CCD benchs}
	\end{center}
    \end{minipage}
\end{figure}

The XMM-Newton observatory, by virtue of its large FOV and high throughput, provides good capabilities for the detection of
low surface brightness emission features from extended and diffuse galactic and extragalactic sources. In order to fully exploit
this, an understanding of the background is required.
There are several different components to the EPIC background:
\begin{itemize}
\item Astrophysical background which is dominated by thermal emission at energies below 1 keV and a power-law at higher energies 
(primarily from unresolved cosmological sources) and sometimes a contribution from solar wind charge exchange. 
This background varies over the sky at lower energies.
\item Particle background, consisting of soft proton flares (with spectral variations from flare to flare) 
and that induced by cosmic rays, both directly by particles penetrating the CCDs and indirectly by fluorescence of instrument and spacecraft materials.
\item Electronic noise consisting of CCD readout noise (mainly below 200 eV) and effects of detector defects such as bad pixels.
\end{itemize}

As part of the EPIC instrument, the EPIC Radiation Monitor (ERM) experiment 
registers the total count rate and basic spectral information of the background radiation impinging on the spacecraft. 
Its main purpose is to supply particle environment information for the safe and correct operation of the EPIC cameras.


\section{\textit{The Reflection Grating Spectrometers (RGSs)}}
\label{sec:RGS}

In two of the three XMM-Newton X-ray telescopes (those having EPIC-MOS cameras),
about half of the X--ray light is utilized by the RGSs.
A detailed description of the instrument is given in \cite{denHerder2001}.

The spectral range covered by the RGS of 5-38 \AA has a high density of emission lines including the K-shell transitions and He-like triplets of light elements, such as C, N, O, Ne, Mg, and Si as well as the L-shell transitions of heavier elements such as Fe and Ni, thus offering a large number of diagnostic tools to investigate the physical conditions and chemical composition of the emitting material.
 
\begin{figure}[th]
	\begin{center}
		
		\includegraphics[scale=1]{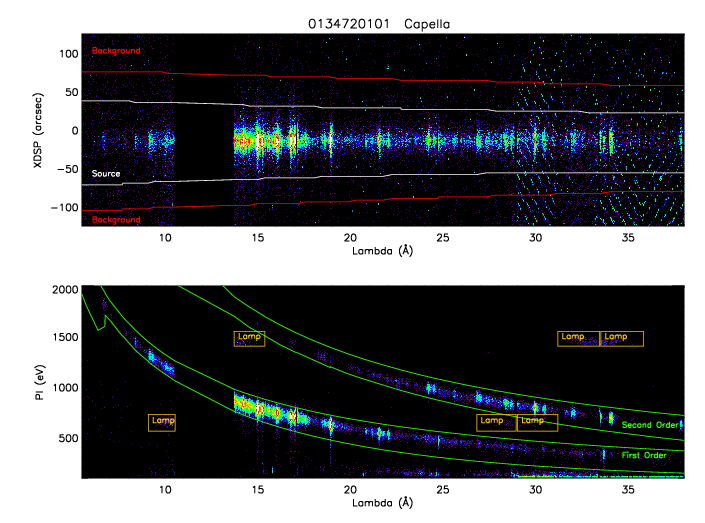}
		
		\caption{\label{Fig:banana}  RGS data for an observation of Capella. The dispersion axis runs horizontally and increases to the right. The top panel shows the image of the dispersed light in the detector. The cross dispersion is along the vertical axis. The bottom panel shows the order selection plane, with the energy (PI), on the ordinate. This also illustrates the mechanism used for separation of first, second and higher grating orders. Standard data extraction regions  are indicated by the curves}
		
	\end{center}
	
\end{figure}

\subsection{\textit{The Instrument}}

Each RGS consists of an array of reflection gratings which diffract X-rays to an array of nine dedicated Charge-Coupled Devices (CCDs), located  along the spectrograph Rowland circle.
Both spectrometers cover the same FOV, with the dispersion direction along the spacecraft Z-axis. The size of the FOV in the cross-dispersion direction is 5$^\prime $.
In the dispersion direction, the aperture of RGS covers the entire FOV of the mirrors, although the effective area decreases significantly for off-axis sources. Each instrument consists of two identical chains with the following units:

\begin{itemize}
	\item The Reflection Grating Array unit (RGA); see Fig.~\ref{Fig:RGA}.
	\item The Focal Plane Camera unit (RFC), including the detectors and the front-end electronics;
           see; Fig.~\ref{Fig:RFC}.
	\item The Analogue Electronic units (RAE) that process data and control the readout sequence of the CCDs.	
	\item Two Digital Electronic units (RDE) that control the instrument.
\end{itemize}

The RGAs consist of an array of reflection gratings (182 for RGS1, 181 for RGS2). Each RGA intercepts 58\%\ of the total light focused by the mirror module. 
The grating plates have mean groove densities of about 645.6 lines mm$^{-1}$. The dispersion of the instrument is a slowly varying function of dispersion angle, approximately 8.3 and 12.7 mm \AA$^{-1}$ at 15 \AA\  in first and second order, respectively.

The dispersion equation for the spectrometer is given by:

\begin{equation}	\label{eq:dispersion}		
	 m\  \lambda = d\ (\cos \beta - \cos \alpha) 
\end{equation}

where $m$ is the spectral order (-1, -2...), d is the groove spacing, $\beta$ is the angle between the outgoing ray and the grating plane, and $\alpha$ is the angle between the incoming ray and the grating plane. The light is primarily diffracted into the "inside" spectral orders, where $m< 0$, so that $\beta > \alpha$.

The RGA diffracts the X-rays to an array of nine MOS back-illuminated CCDs. Each has 102 4$\times$  768 pixels, half exposed to the sky and half used as a storage area. During readout, 3 $\times$ 3 pixel on-chip binning is performed, leading to a bin size of (81$\mu$m)$^2$, which is sufficient to fully sample the line spread function, reducing the readout time and the readout noise. In the dispersion direction one bin corresponds to about 7, 10, and 14 m\AA\ in first order and about 4, 6, and 10 m\AA\ in second order for wavelengths of 5, 15 and 38 \AA, respectively. The size of one bin projected onto the sky is about 2.5'' in the cross-dispersion direction and roughly 3, 5 and 7'' and 4, 6, and 9'' in the dispersion direction at 5, 15 and 38 \AA\ in first and second order, respectively.
  
\begin{figure}
	\begin{center}
		\includegraphics[angle=90,scale=0.4]{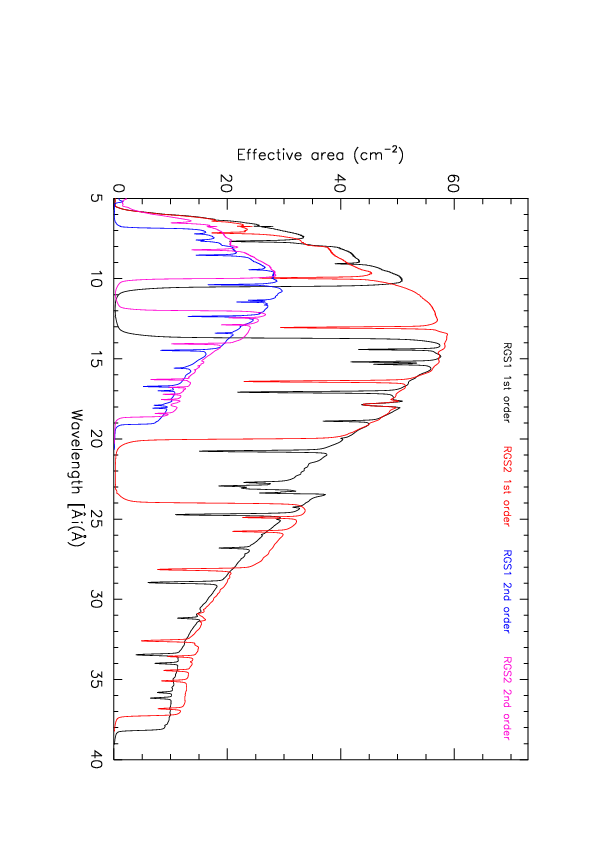}
				\caption{\label{Fig:area} Example of the RGS effective area for a recent observation. Clearly visible are
			the gaps due to the non-working CCDs and, in RGS1, the prominent instrumental O edge near 23 \AA}
		\end{center}
	\end{figure}

After the first week of operations, an electronic component in the clock driver of CCD4 in RGS2 failed, affecting the wavelength range from 20.0 to 24.1 \AA. A similar problem occurred in early September 2000 with CCD7 of RGS1 covering 10.6 to 13.8 \AA. The total effective area is thus reduced by a factor of two in these wavelength bands (see Fig.~\ref{Fig:area}).

In 2007, the readout method in RGS2 was changed from double-node, in which data from the two halves of the chips are retrieved separately, to single-node, in which data from the whole chip are read out through a single amplifier. Hence, RGS2 frame times are twice as long as those from RGS1.

The standard mode of operation of the RGS instrument is called "Spectroscopy". It consists of a two-dimensional readout of one or more CCDs over the full energy range. The accumulation time when reading the eight functional CCDs is 4.8 s for RGS1 and 9.6 s for RGS2.

To mitigate the effects of pile-up, very bright sources can be observed in the RGS "Small Window" mode. In this mode, only the central 32 of the 128 CCD rows in the cross-dispersion direction are read. The CCD readout time is therefore decreased by a factor 4 compared to Spectroscopy mode. It can be further decreased by reading only a subset of the CCDs.

\subsection{\textit{Scientific Performance}}

A consequence of the diffraction equation (\ref{eq:dispersion}) is that orders overlap on the CCD detectors of the RFC. Separation of the spectral orders is achieved by using the intrinsic energy resolution of the CCDs, which is about 160 eV FWHM at 2 keV. The dispersion of a spectrum on an RFC array is shown in the bottom panel of Fig.~\ref{Fig:banana}. First and second orders are very prominent and are clearly separated in the vertical direction (i.e., in CCD energy, or PI, space). Photons of higher orders are also visible for brighter sources. The calibration sources can also be seen in the bottom panel as short horizontal features
(Fig.~\ref{Fig:banana}).

\begin{table}   
\begin{center}
\begin{tabular}{| l l | c c c | c c c |} \hline
	
&                        & \multicolumn{3}{| c |}{RGS1} & \multicolumn{3}{| c |}{RGS2}  \\

\hline
&                        & 10 \AA\ & 15 \AA\ & 35 \AA\  & 10 \AA\ & 15 \AA\  & 35 \AA\ \\ \hline \hline

\hline
Effective area (cm$^2$)  & 1$^{st}$ order  & 51    & 61   & 21    & 53    & 68   & 25 \\
						 & 2$^{nd}$ order  & 29    & 15   & --    & 31    & 19   & -- \\
\hline
Resolution (km s$^{-1}$) & 1$^{st}$ order  & 1700  & 1200 &  600  & 1900  & 1400 & 700 \\
						 & 2$^{nd}$ order  & 1000  &  700 &  --   & 1200  &  800 & --  \\
\hline
Wavelength range         & 1$^{st}$ order  & \multicolumn{6}{| c |}{5 -- 38 \AA\ (0.35 -- 2.5 keV)} \\
						 & 2$^{nd}$ order  & \multicolumn{6}{| c |}{5 -- 20 \AA\ (0.62 -- 2.5 keV)} \\
\hline
Wavelength accuracy      & 1$^{st}$ order  & \multicolumn{3}{| c |}{$\pm$5 m\AA} & \multicolumn{3}{| c |}{$\pm$6 m\AA}\\
					     & 2$^{nd}$ order  & \multicolumn{3}{| c |}{$\pm$5 m\AA} & \multicolumn{3}{| c |}{$\pm$5 m\AA}\\
\hline
\end{tabular}
\caption{{RGS In-orbit Performance}}
\label{tab:rgskey}
\end{center}	
\end{table}

A summary of the RGSs' performance is given in Tab \ref{tab:rgskey}.
A complete overview of the performance of the RGS, instrumental details and calibration procedures can be found in \cite{deVries2015}. 

\medskip

The calibration of the RGS effective area is based on a combination of ground measurements and in-flight observations (Fig. \ref{Fig:area}). Empirical corrections have been introduced along the years, the first one based on the assumed power law form of blazar spectra, followed by the recognition of wavelength-dependent sensitivity changes  consistent with a buildup of hydrocarbon contamination on the CCD surface. 
There are indications of a continuous decrease in effective area over the last years, in both instruments and affecting  most of the spectral range. This decrease cannot be explained by only contamination by hydrocarbons. Its origin is not understood. This calibration takes into account this effect, following an empirical algorithm.
The first-order effective area peaks around 15 \AA\ (0.83 keV) at about 120 cm$^2$ for the two spectrometers.

The wavelength scale is determined by the geometry of the various instrument components. The original pre-flight calibration kept the wavelength scale accuracy well within specification. Nevertheless, it has been improved by taking into account some systematic effects. With these corrections, the accuracy of the wavelength scale is now of order of 6 m\AA.

The observed line shape is well represented by the model. The empirically determined width of strong emission lines is a slowly varying function of wavelength in both instruments, with a mean FWHM of about 70 m\AA\ in first order and 50 m\AA\ in second order, giving a spectral resolution that increases with wavelength. It is estimated that an observed line broadening of more than 10\%\ of the FWHM can be considered to be significant for strong lines.

The current status of the instruments and the calibration can be found in the "XMM-Newton Users Handbook" (\cite{Ebrero2021}) and in the document "Status of the RGS Calibration" \cite{Gonzalez2021}, both available at the XMM-Newton website.


\begin{figure}[h]
\begin{center}
\includegraphics[angle=0,width=0.95\textwidth]{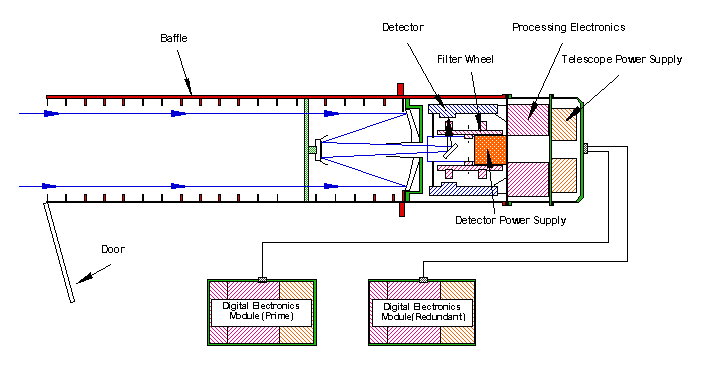}
\caption{Schematic of the Optical Monitor}
\label{fig:om_schematic}
\end{center}
\end{figure}

\section{\textit{Optical Monitor (OM)}}
\label{sec:OM}

The Optical Monitor (OM) provides simultaneous optical/UV coverage of
sources in the EPIC field of view, extending the wavelength range of the
mission and enhancing its scientific return.

The photon-counting nature of the instrument and the low in-space
background mean it is highly sensitive for the detection of faint sources,
despite its small size, being able to reach about magnitude 22 (5$\sigma$
detection) in the B filter in 5 ks of exposure (with maximum depth in the
White filter). The provision of UV and optical grisms permits low-resolution
spectroscopic analyses, while the fast mode timing options allow detailed
studies of temporal variability.

\subsection{\textit{The Instrument}}

\begin{figure}[t]
  \begin{minipage}[t]{5.5cm}
    	\begin{center}
	\includegraphics[angle=0,scale=0.5]{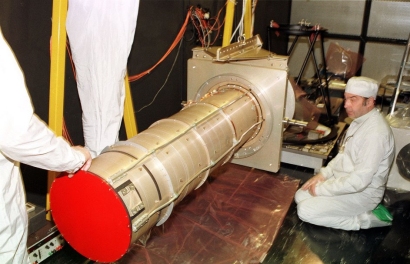}
	\caption{\label{Fig:OM1}
          The Optical Monitor at Mullard Space Science Laboratory, UK}	 
	\end{center}      
    \end{minipage}
  \hfill
  \begin{minipage}[t]{5.5cm}
    \begin{center}
		\includegraphics[angle=0,scale=0.8]{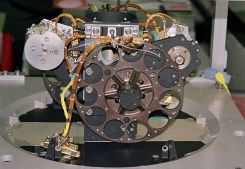}
		\caption{\label{Fig:OM2}
                  One of the two filter wheels positioned in front of the OM's CCD detectors}
	\end{center}
    \end{minipage}
\end{figure}

The OM is a 2 m-long, 30 cm diameter telescope of Ritchey-Chretien design,
with a focal length of 3.8 m (f/12.7). After passing through the primary
mirror hole, the light beam impinges on a rotatable 45$^\circ$ flat that
deflects it to one of two redundant detector chains. Each chain comprises
a filter wheel containing 11 apertures (V (500-600 nm), B (380-500 nm),
U (300-400 nm), UVW1 (220-400 nm), UVM2 (200-280 nm), UVW2 (180-260 nm) and
White-light (180-700 nm) broadband filters, visible (290-600 nm) and
UV (180-360 nm) grisms for dispersive (resolving power
($\lambda/{\Delta{\lambda}}$) $\sim$ 180) spectroscopy, a
magnifier\footnote{not available for use}, and a mirror acting as a blocking
filter).
A schematic of the OM is shown in Fig.~\ref{fig:om_schematic},
while the photographs in Figs. \ref{Fig:OM1}   and  \ref{Fig:OM2}
show the telescope and the filter wheel assembly, respectively.

The detector, located behind the filter wheel in each chain, is a
Micro Channel Plate (MCP)-intensified CCD (MIC), comprising a S20
photocathode, a pair of MCPs, a P-46 phosphor, a fiber taper and
a CCD. An electron liberated from the photocathode by an incident
sky photon drifts to the upper MCP where a potential accelerates
it along a pore, creating a cascade of electrons by collisions with
the pore walls. On passing through the second MCP, this charge cloud is 
amplified to around 5 $\times$ $10^5$ - $10^6$ electrons, and these impinge
on the phosphor, resulting in a burst of photons. This photon burst,
spatially localized by the MCP arrangement, then traverses the fibre taper,
onto the CCD, which has 256 $\times$ 256 light-facing pixels (each 4$\times$ 4 arcsecs
on the sky). The footprint
of the photon burst at the CCD covers about 3 $\times$ 3 CCD pixels. On readout, an
onboard algorithm then centroids each footprint to 1/8 of a CCD pixel,
creating an effective image of 2048 $\times$ 2048 image pixels (maximum resolution,
0.5 $\times$ 0.5 arcsec pixels on the sky). The CCD is read out about 90 times (frames)/s (for the
full field). Thus, each sky photon incident on the photocathode
that yields a footprint at the CCD within the frame is subject to
onboard validation thresholds, recorded as an incident event by the
onboard data processing system.

The OM has two main modes of operation:
Imaging mode, where events from each frame are accumulated into a single image
covering the total exposure time,
and Fast mode, where, for a small 11'' $\times$  11'' window, each event is
time-tagged, yielding an event stream. In Imaging mode, the observer has,
subject to telemetry-related constraints, significant freedom to define 
window(s) for optimum sky coverage for their science goals. This may involve coverage
of the whole field, at the expense of longer instrument overheads, or coverage
of more localized areas of the field via up to five smaller windows.
Grism data is taken either in a Full-field mode, potentially yielding spectra
from all sufficiently bright objects in the field, or with a narrow,
predefined window, designed to concentrate on a specific target observed at
the boresight. In Fast mode only two Fast mode windows are allowed, though
normally these can be used in conjunction with image mode windows. The
highest time resolution in fast mode is 0.5 s. 

\begin{figure}[h]
\begin{center}
\includegraphics[angle=0,width=0.95\textwidth]{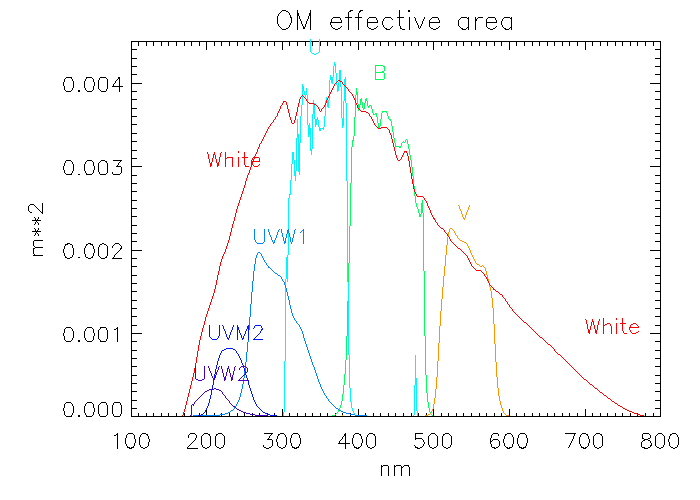}
  \caption{The effective area curves for each of the OM photometric filters,
  essentially at the start of XMM-Newton mission operations (2000)}
\label{fig:effareas}
\end{center}
\end{figure}

Three important consequences of the OM design on its output data are as follows:
(1) When two or more photon bursts arrive at the CCD within the same
readout frame and their footprints spatially overlap, the probability
of which increases with source count rate and/or longer frame times,
they may not be distinguished and so be recorded as a single event.
This effect is referred to as coincidence loss and is similar to the
pile-up effect in the EPIC cameras. (2) The instrument design,
particularly the fiber taper, results in a distortion of the imaged
field compared to the real sky. (3) For speed, the onboard algorithm
exploits a lookup table to centroid the count distribution in
each 3 $\times$ 3 CCD-pixel footprint, a simplification that results in a
so-called ``modulo-8'' pattern appearing on a scale of 8 pixels in the
2048 $\times$ 2048 output image. These effects are generally corrected for
through software tools in the XMM-Newton Science Analysis Software (SAS).
Some OM observations also contain low-intensity, diffuse light features, arising from
reflections from the back side of the detector entrance window and/or from a chamfer around it.

The OM, being a photon-counting instrument, has high sensitivity and, with the
low background (dominated by zodiacal light), it can reach stars as faint as
about V=21 (for a 5$\sigma$ detection of an A0 star in the B filter) in 1000 s. 
On the other hand, the photocathode can be damaged by high incident photon
rates, and this places limits on the brightest sources that the OM can be
exposed to. In practice, the limiting magnitude is around V = 7.3 for an
A0 star. A more complete description of the OM instrument is given in \cite{Mason2001}.

\begin{figure}[h]
\begin{center}
\includegraphics[angle=0,width=0.95\textwidth]{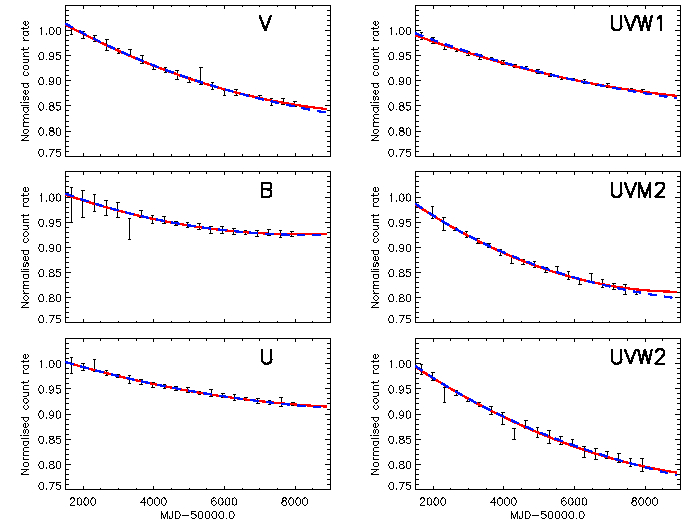}
\caption{Normalized, observed count rates of constant sources in
  the OM SUSS4 catalogue, binned into 20 time bins (black error bars), for
  each OM filter. The solid red and dashed blue curves represent
  the most recent and previously fitted decline trends, respectively,
  highlighting the ubiquitous flattening of the decline }
\label{fig:TDS}
\end{center}
\end{figure}

For observations not performed in Full-Frame mode, the OM performs a short
(20 s) field-acquisition exposure at the start of each observation (\cite{Talavera2011}).
This
enables a number of pre-specified stars to be recognized in the exposure
by the onboard software and the observed and predicted positions compared
to measure shifts due to uncertainties in the spacecraft pointing. Consequently,
the chosen OM science observation windows can be adjusted in position to ensure
the sky coverage is optimal for the observer's science. This is especially important
for accurately positioning the small Fast mode windows, when used, to ensure the
target is well centered in the window.

In addition, the star positions are also monitored every 20 s, permitting
the tracking of any spacecraft drift. This tracking information is used by the
onboard software, to relocate events to the correct position in the accumulating
image for Image mode data (referred to as "shift and add"). For Fast mode data,
this tracking information is not applied onboard but is used for the same purpose
in downstream data reduction performed by the SAS.

\subsection{\textit{Scientific Performance}}

The OM has proven to be a very stable instrument. Nevertheless, it has
experienced some spatial and temporal changes in sensitivity over the long
baseline of the XMM-Newton mission, both expected and unexpected  (\cite{OMcalstatus2020}).

Of particular note are the effective areas. These were determined
soon after launch, for each photometric filter, based on
measurements of spectrophotometric standard stars, alongside the
conversions from count rate to absolute flux and the equivalent
magnitude zero-point determinations. The effective area for each filter, 
initially modelled from pre-launch information of the optical
components (e.g., mirror area, reflectivities, filter transmissions), and the
quantum efficiency of the photocathode were subsequently adjusted, in-flight,
to match the observed count rates of standard stars. It was found that all filters
showed reduced sensitivity (from 16\% to 56\% residual throughput) compared
to pre-launch expectations, with the UVM2 and especially UVW2 filters most affected.
The reduction in sensitivity has been adequately modelled by absorption
due to a molecular contaminant layer somewhere in the OM system (\cite{Kirsch2005}). The
effective areas of the photometric filters, essentially at launch,
are shown in Fig.~\ref{fig:effareas}.

Subsequently, anticipated aging of the detector and, likely, some
further contaminant growth have resulted in a gradual decline in
sensitivity since launch. This decline, known as the time-dependent
sensitivity (TDS) degradation, is filter (wavelength) dependent. It is
monitored and characterized via analysis of data from the OM Serendipitous
UV Sky Survey (SUSS) catalogues (\cite{Page2012}) and is routinely verified
via observations of spectrophotometric standard stars. The most recent
TDS trends for the narrowband filters are shown in Fig.~\ref{fig:TDS}.
The decline in sensitivity ranges from around 7\% in the B filter to
about 22\% in the UVW2 filter. These curves are used to correct the
observed count rates of sources at any epoch within the mission baseline
to the rate expected at the start of the mission. That rate can then
be converted to absolute photometric values via the at-launch flux and
zero-point conversions.


\section{\textit{Organization of the XMM-Newton Ground Segment}}
\label{sec:ground}

The Mission Operations Centre (MOC) at the European Space Operations Centre (ESOC), Darmstadt, Germany, controls the spacecraft 24 h per day, all year round, using, as main ground stations,  Kourou (French Guiana) and Sanitago (Chile) and various other additional stations in South America and Australia.
The MOC is responsible for the maintenance and operations of the spacecraft and  the required ground infrastructure. As XMM-Newton has no onboard data storage capacity, all data are immediately down-linked to the ground in real time. Since 2018 XMM-Newton has been operated together with Gaia and INTEGRAL.

The Science Operations Centre (SOC) at the European Space Astronomy Centre (ESAC), Villanueva de la Ca{\~n}ada, Madrid, Spain, is responsible for science operations and for supporting the scientific community. 
The SOC handles Announcements of Opportunity and proposals, including technical evaluation and OTAC support, as well as the subsequent planning of observations, including instrument handling, calibration observations, and Targets of Opportunity. 
Data from these observations are processed from raw telemetry to standard data products at the SOC, before being ingested into the XMM-Newton Science Archive (XSA) and distributed to the users. 
A Quick-Look Analysis of data and anomaly monitoring of the instruments are part of this process.
The SOC also takes a leading role in the continuous calibration of the instruments and the provision of scientific analysis software (SAS) to the users together with experts from the XMM-Newton community.

The Survey Science Centre (SSC) \cite{Watson2001}, a consortium of 10 institutes in the ESA community, is responsible for the compilation of the XMM-Newton Serendipitous Source Catalogue, the follow-up/identification program for the XMM-Newton serendipitous X-ray sky survey, support to pipeline processing at the SOC, and development of parts of the scientific analysis software.

NASA provides a Guest Observer Facility (GOF) at the Goddard Space Flight Center (GSFC), Greenbelt, Maryland, USA. The GOF supports the usage of XMM-Newton by the scientific community in the USA.
It distributes the XMM-Newton data to US users and contributes to the SAS development. 
The GOF is responsible for the organization of the Guest Observer (GO) program funded by NASA.


\section{\textit{Observing with XMM-Newton}}
\label{sec:obs}

All the scientific instruments onboard XMM-Newton can be operated
independently and can obtain science data simultaneously, if
operational constraints permit. These constraints are imposed to
preserve the safety of the instruments as well as to achieve the
conditions for an optimal calibration of the data. The combination of
the orbit of the spacecraft with limits in the instruments' operational
parameters (mostly temperatures but also radiation dose) results in
observation constraints related to the orientation of the spacecraft
with respect to the Sun, Earth and Moon and to the position of the
spacecraft in the Earth's magnetosphere.

XMM-Newton was launched in December 1999
into a highly elliptical orbit, with a high inclination with
respect to the Equator and with an apogee height of 115,000 km
in the Northern hemisphere and a perigee height of 6000 km in the
Southern hemisphere.

Due to several perturbations, the orbit of XMM-Newton evolves with
time.  An orbit correction maneuver was performed in February 2003 to
ensure full ground station coverage during the entire science
period. But the evolution of the orbit has changed the fraction of the
sky visible to science instruments and the effective available science
time per orbit along the mission lifetime.

The spacecraft has no capacity for data or commanding storage onboard
so it requires continuous contact with the ground for science operations.
The operations are conducted from the MOC through ground stations in Kourou, Santiago de Chile and
Yatharagga (and in Perth and New Norcia during the early years of the mission
and occasionally in Madrid).

\begin{figure}
  \includegraphics[width=200bp]{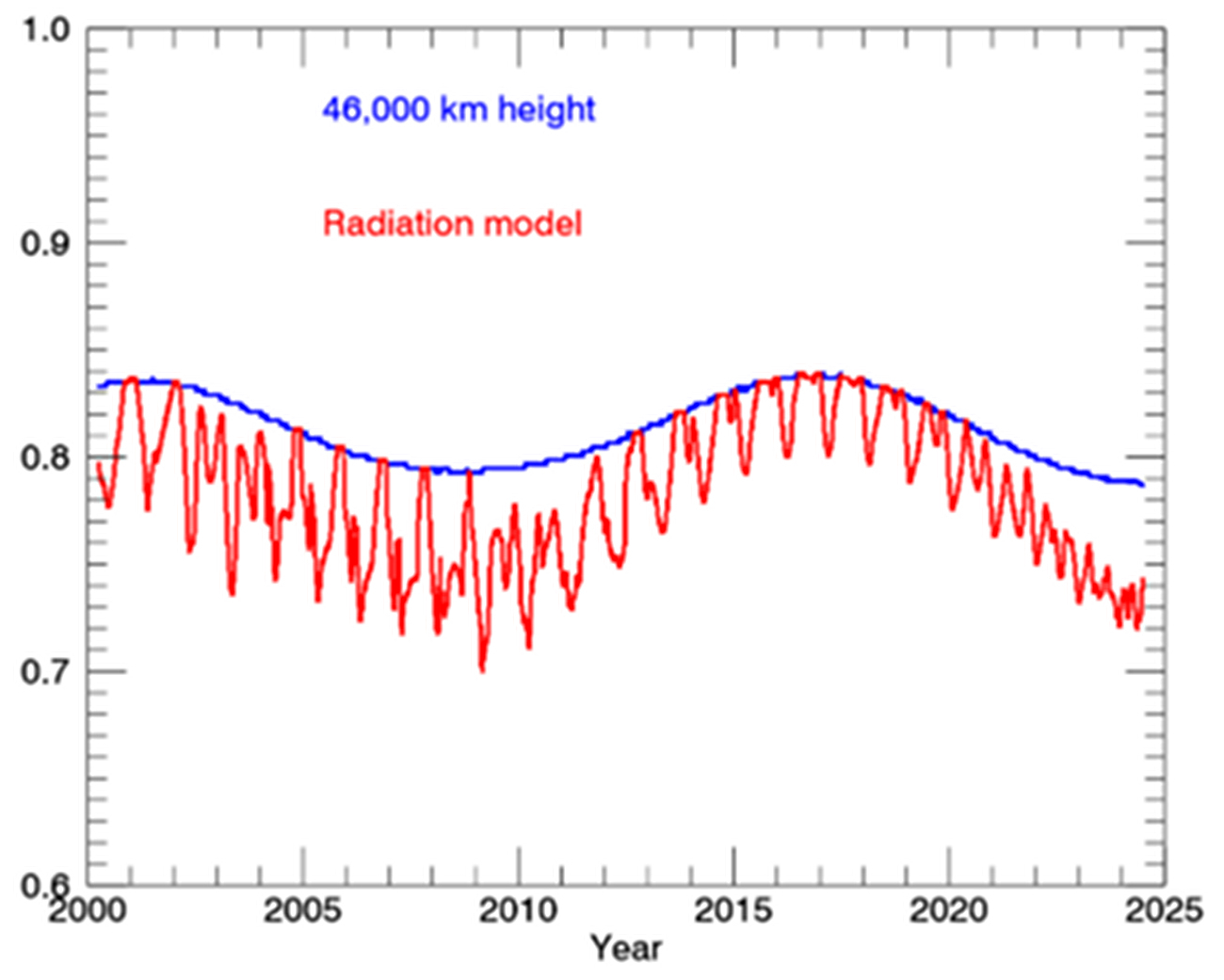}
  \caption{The evolution of the XMM-Newton orbit during the mission has changed
    the spacecraft velocity near perigee passage and the 
    orientation with respect to Earth magnetosphere. As a result the fraction
    of time available to science has changed with a long-term trend superposed on a
    seasonal modulation}
  \label{fig:P1}
\end{figure}

The EPIC and RGS CCD detectors are sensitive to both X-ray and
optical radiation as well as particles. Electromagnetic
radiation can affect the scientific analysis of the data collected,
but protons striking the detectors can permanently damage the surface
of the CCDs. In order to protect the EPIC cameras, their filter wheel
is moved into the closed position during intervals of high particle
radiation. The RGS spectrographs do not have a similar filter protection
so the instruments are placed into a special configuration with minimal
equipment switched on during high radiation intervals.

Before launch, it was expected that the radiation environment above 46,000 km from the
Earth was safe for the mission. This meant that
$\sim$143 ks of the $\sim$173 ks ($\sim$48 h) orbital period
could be devoted to science operations at the beginning of the mission.
However, strong fluctuations and variability of the particle background in the
cameras were one of the main surprises and concerns following launch. 
An ad hoc model for radiation belts around the Earth was developed in
order to predict the time window within every revolution when science observations
could be safely conducted
\cite{Casale2004}; see Fig.~\ref{fig:P1}.
According to the model and the expected orbital evolution, the science
window will be as short as $\sim$130 ks by 2025.

Many parameters in the EPIC cameras and RGS spectrographs are strongly
dependent on the temperatures at which the instruments are operated. In order
to guarantee a consistent calibration for all observations the
operations are designed to maintain the temperatures of the
science payload within a narrow range. Since the main source of heating
is illumination by the Sun, strong constraints on temperatures
are translated into a strong constraint on the orientation of the spacecraft
with respect to the Sun. The solar aspect angle (angle between pointing
direction and Sun direction) must be within 70$^{\circ}$\ and 110$^{\circ}$\ at all times
to assure thermal stability and sufficient power from the solar array.
This means that $\sim$\,65\%\ of the sky is inhibited in every single XMM-Newton revolution.

Other celestial constraints are unrelated to temperature or power
stability, but to potential electromagnetic radiation damage of the
OM.  The main sources of dangerous light emission,
away from the Sun, are the Earth and the Moon. The Earth limb avoidance
angle is 42.5$^{\circ}$, and the Moon limb avoidance angle is 22$^{\circ}$,
which is increased to 35$^{\circ}$\ during eclipse seasons.

\begin{figure}
  \begin{minipage}[t]{5.5cm}
    	\begin{center}
		\includegraphics[angle=0,scale=0.065]{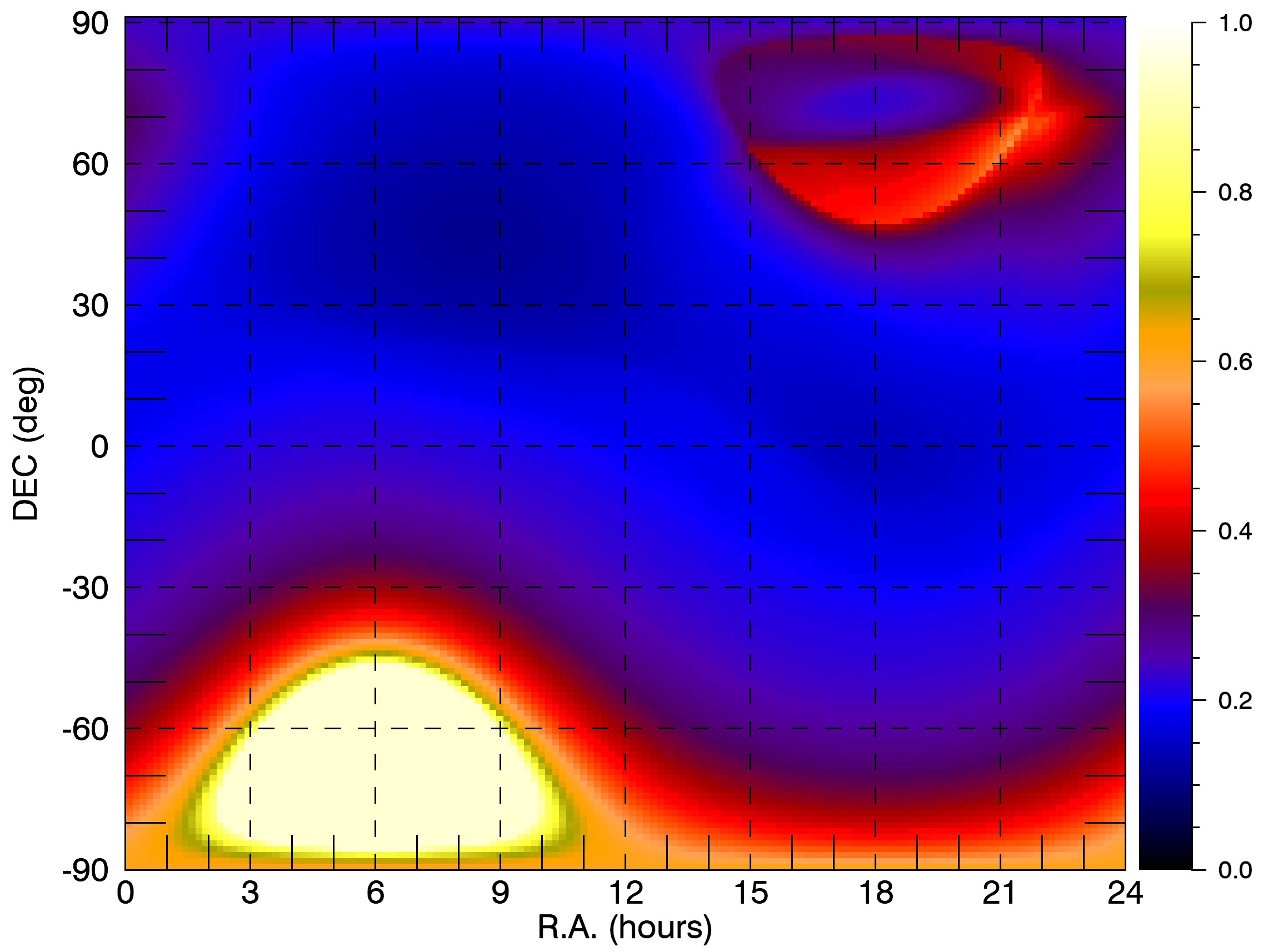}
	\end{center}
    \end{minipage}
  \hfill
  \begin{minipage}[t]{5.5cm}
    	\begin{center}
	\includegraphics[angle=0,scale=0.065]{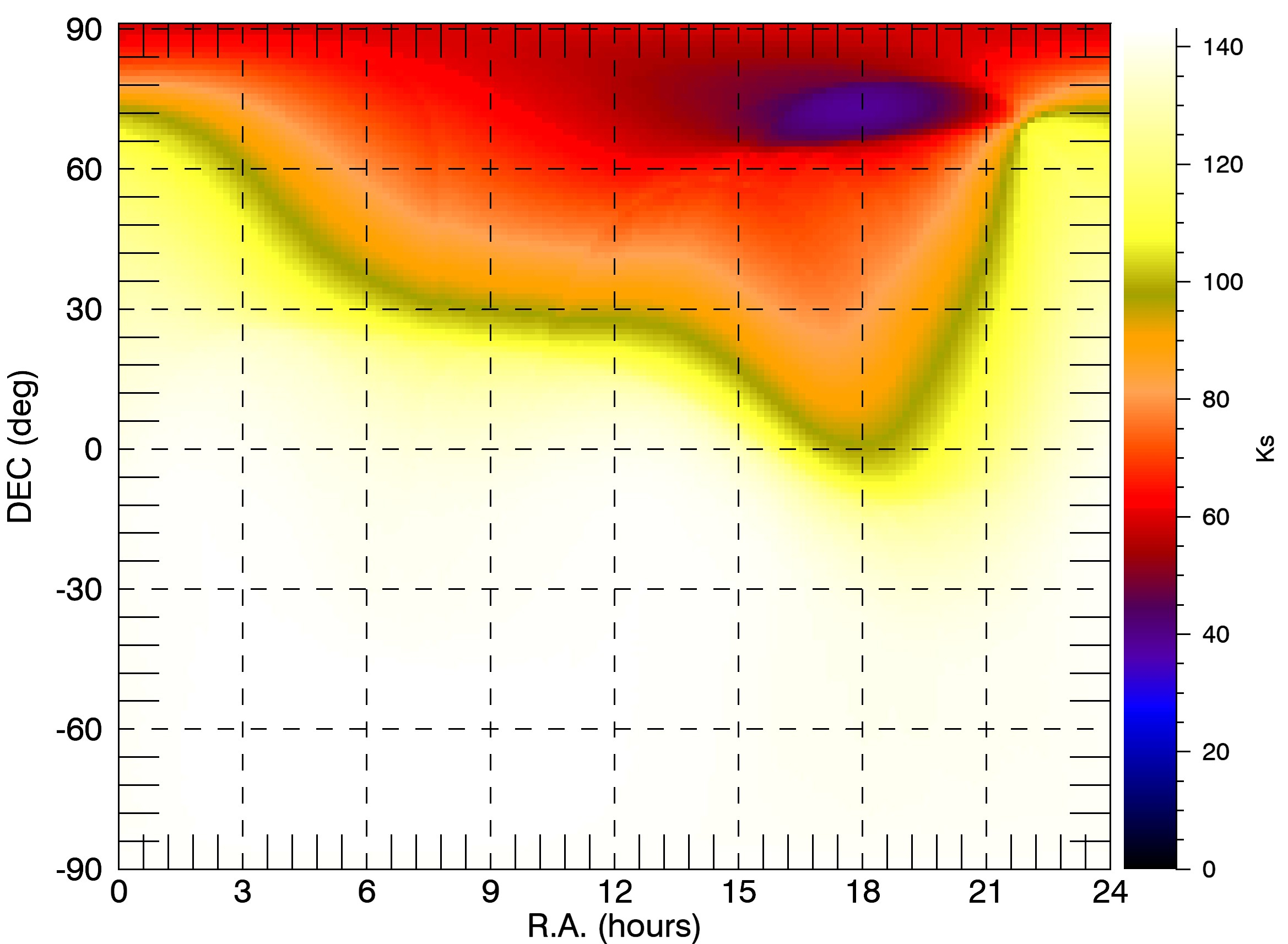}	 
	\end{center}
  \end{minipage}
   \caption{The high inclination of the orbit and the high latitude
    of the perigee have made the South Ecliptic Pole the region of the
    sky with the best accessibility to XMM-Newton along its lifetime.
    By contrast, the visibility around the North Ecliptic Pole is
    constrained by the Earth in most of the revolutions.
    The image shows the fraction of visible orbits along 21 years (right)
and the average maximum visibility in 21 year (left)
   }
    \label{fig:H1}
\end{figure}

As in the case of the radiation constraints, the evolution of the
XMM-Newton orbit with respect to the ecliptic has had consequences
on the fraction of the sky available at any time and on the
evolution of the visibility in certain areas of the sky; see Fig. \ref{fig:H1}.

The constraint outlined above apply to any type of observation,
independent of the configuration of the instruments. There are a
number of constraints that apply only to OM exposures and some science
exposures cannot be performed using the OM, but exposures with the X-ray
instruments are permitted. These OM constraints refer to the presence
of nearby bright celestial sources that may damage the detector. In
addition, OM exposures are forbidden near the following solar system
objects (Table \ref{tab:omav}):

\begin{table}   
\begin{center}
\begin{tabular}{|l|c|} \hline
 Planet  & Avoidance angle \\ \hline \hline
 Mars    & 3.5$^{\circ}$ \\
 Jupiter & 4.5$^{\circ}$ \\
 Saturn  & 2$^{\circ}$ \\
 Uranus  & 0.25$^{\circ}$ \\
 Neptune & 0.25$^{\circ}$ \\ \hline
\end{tabular}
	\caption{{OM avoidance angles}}
	\label{tab:omav}
\end{center}
\end{table}


\section{\textit{Scientific Data and Analysis}}
\label{sec:data}

XMM-Newton reformatted telemetry is organized in Observation/Slew Data
Files (ODF/SDF). Most of the ODF/SDF components have a FITS format. 
An ASCII summary file provides the astronomers with general information 
on the observation and an index of the files contained in the ODF.

The Science Analysis System (SAS) is the software established to reduce 
and analyze XMM-Newton science data. It consists of two main blocks:
\begin{itemize}
\item reduction pipelines, which apply the calibrations to the ODF and the 
SDF science files and produce calibrated and concatenated event lists 
for the X-ray cameras, flat-fielded and calibrated OM sky images, 
source lists, and time series.
\item a set of file manipulation tools, which include the
extraction of spectra, light curves, and (pseudo-)images and the 
generation of source lists, as well as the generation of auxiliary files 
such as appropriate instrument response matrices.
\end{itemize}

The SAS reduction pipeline (PPS) is run on all XMM-Newton datasets.
Each PPS dataset is manually screened to verify its scientific quality 
and identify potential processing problems.
The PPS output (\cite{SSCProducts}) includes a wide range of
top-level scientific products, such as X-ray camera event lists, 
source lists, multi-band images, background-subtracted spectra, 
and light curves for sufficiently bright individual sources, as well as 
the results of a cross-correlation with a wide sample of source 
catalogues and with the matching ROSAT field.

All the XMM-Newton calibration data are organized in a Current
Calibration File (CCF). Summary documents, containing an overview of the current calibration 
status and associated systematic uncertainties, are available from 
the XMM-Newton Calibration Portal.

The XMM-Newton Science Archive (XSA) content is regularly updated 
with all the newly generated ODF, SDF, and PPS products, with updated versions 
of the catalogues of EPIC sources, OM sources, and Slew Survey 
sources, and with ancillary info like associated proposal 
abstract and publications. 
On-the-fly data analysis and processing can be performed from the
XSA using the SAS without the 
need of downloading data or software.

With all the sources serendipitously detected in the EPIC FOV of
XMM-Newton public observations, the SSC compiles and regularly updates 
the XMM-Newton EPIC source catalogue. 
At the time of writing, the SSC has created four catalogue
generations, with 4XMM
being the latest, with a few incremental versions for each of them, 
leading so far to a total of 11 catalogue data releases, 11DR.  
4XMM-DR11 (\cite{Webb2020}) was released in August 2021 and 
contains 787,963 ``clean'' detections corresponding to 602,543 
unique sources covering 1239 sq. deg. 
In addition, the 4XMM-DR11s catalogue of serendipitous sources 
detected from stacked data from 
overlapping XMM-Newton observations  (\cite{Traulsen2020}) 
contains 358,809 unique sources 
of which 275,440 were multiply observed covering 350 sq. deg. 
The data acquired during satellite slews are used to build the
XMM-Newton Slew Survey Catalogue, XMMSL2 (\cite{Saxton2008}).
The current version contains 55,969 clean detections covering an area of
650,000 sq. deg. 
The OM team, under the auspices of the SSC, produces and regularly updates a
catalogue of sources detected by the Optical Monitor.
The 5th version of the XMM-Newton OM Serendipitous Ultraviolet Source
Survey Catalogue, XMM-SUSS5,
(\cite{Page2012}) contains 8,863,922 detections of 5,965,434 sources. 
Specific queries to all catalogues can be made using the XMM-Newton
Science Archive.


\begin{figure}
	\begin{center}
		\includegraphics[angle=0,scale=0.4]{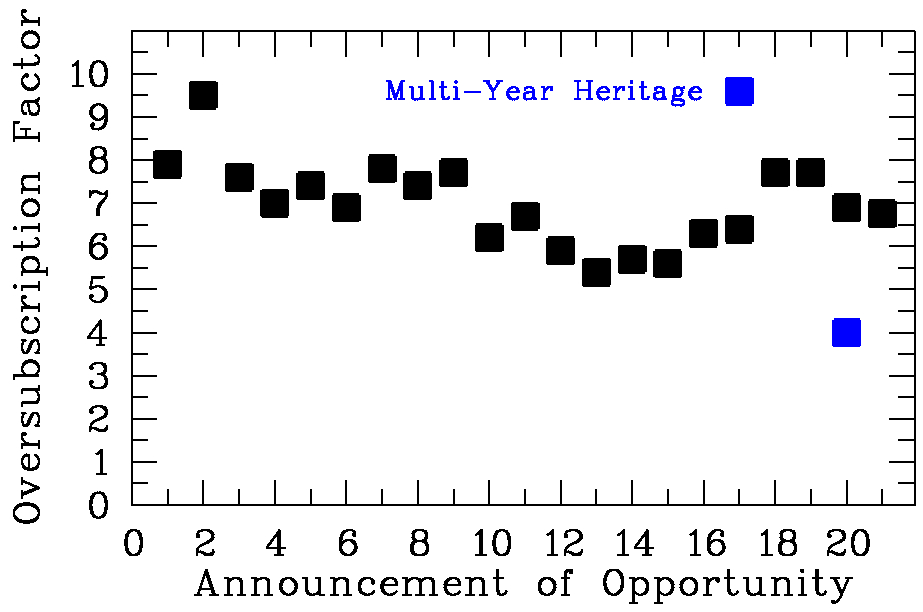}
		\caption{\label{Fig:over}
The over-subscription factors, or requested versus available 
observing time, for the first 21 XMM-Newton 
Announcements of Opportunity (AOs) are shown in black.
The two blue symbols show the over-subscription of the 
Multi-Year-Heritage programs}
	\end{center}
\end{figure}

A database of Upper Limits across the FOV of all
public XMM-Newton pointed and slew observations
has been built with new observations being added as they
become available (\cite{Ruiz2021}). The database is searchable from the XSA interface.

\section{\textit{Scientific Strategy and  Impact}}
\label{sec:stra}

XMM-Newton observing time is made available worldwide via Announcements of Opportunity (AOs). 
The AOs open in the second half of August each year and the results are 
publiciszd in early December.
All the AOs were highly over-subscribed, typically by a factor 6 to 7; see
Fig.~\ref{Fig:over}. The proposals are peer-reviewed by panels composed 
of scientists located worldwide.

The XMM-Newton observing strategy was discussed with the 
community at large at two workshops:
"XMM-Newton: The next Decade" in 2007 \cite{Schartel2008} and 2016 \cite{Schartel2017a}  \cite{Schartel2017b}.
The unique capabilities of the instruments and the long mean observing
time (30 ks) in combination with the possibility of long uninterrupted
observations, foster XMM-Newton's potential for transformative 
science. Here transformative science is understood to be scientific 
results which lead to radically restructuring the scientific 
understanding or as observational confirmation of
central predictions of astrophysical and cosmological theory and modelling.
Examples include:
(1) the non-detection or weak detection of cooling flows in the galaxy 
clusters Abell 1835 \cite{Peterson2001}, Abell 1795 \cite{Tamura2001} and S{\'e}rsic 159-03 \cite{Kaastra2001}  which led to the concept of the coupling of the cosmic evolution of supermassive 
black holes with that of galaxies and 
clusters of galaxies via feedback. This meant that two object classes which were considered
to be completely independent before were from then on understood 
to undergo a strongly coupled evolution;
(2) the detection of transitional millisecond pulsars \cite{Papitto2013},
which confirmed the transition of accretion-powered to 
rotation-powered emission modes in pulsars;
(3) the detection of low magnetic field magnetars  \cite{Rea2010} \cite{Tiengo2013}  which
changed our understanding of the magnetic fields
which cause the short X/gamma ray bursts in repeaters,
(4) the identification of neutron stars within ultra-luminous X-ray sources
\cite{Fuerst2016} \cite{Israel2017}. This changed the understanding of the nature of this 
source class and allowed the study of super-Eddington accretion  \cite{Pinto2016}
(5) the determination of the mass, spin and X-ray corona size 
of supermassive black holes \cite{Fabian2009} \cite{Risaliti2013} \cite{Parker2017} \cite{Alston2020} \cite{Wilkins2021} which quantitatively
describe the inner geometry of AGNs,
(6) the study of tidal disruption events
\cite{Reis2012} \cite{Miller2015}  \cite{Kara2016}  \cite{Lin2017} \cite{Lin2018} \cite{Pasham2019} \cite{Shu2020} which shed light on the details of the accretion 
process and jet launching,
(7) the detection of the warm-hot intergalactic medium \cite{Nicastro2018}, which confirmed cosmic simulations by the detection of the missing baryons.
Further examples of transformative science resulting from
XMM-Newton and Chandra observations are given in two
{\it Nature} review articles by Santos-Le{\'o} et al. \cite{SantosLeo2009} and by Wilkes  et al. \cite{Wilkes2021}.

An analysis of the science results and the discussion of the 2017 workshop showed the increased importance 
of Target of Opportunity observations, large and very large programs, and observations 
joint with other facilities for transformative science.

\begin{figure}
  \begin{minipage}[t]{5.5cm}
    	\begin{center}
		\includegraphics[angle=0,scale=0.235]{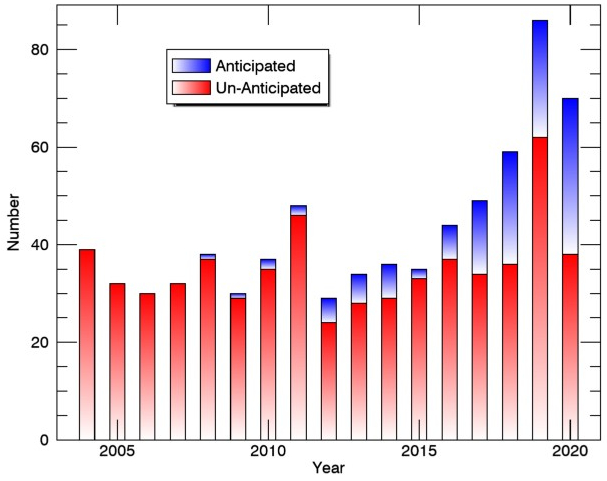}
		\caption{\label{Fig:too}
 Performed Target of Opportunity observations 
}
	\end{center}
    \end{minipage}
  \hfill
  \begin{minipage}[t]{5.5cm}
    	\begin{center}
	\includegraphics[angle=0,scale=0.2]{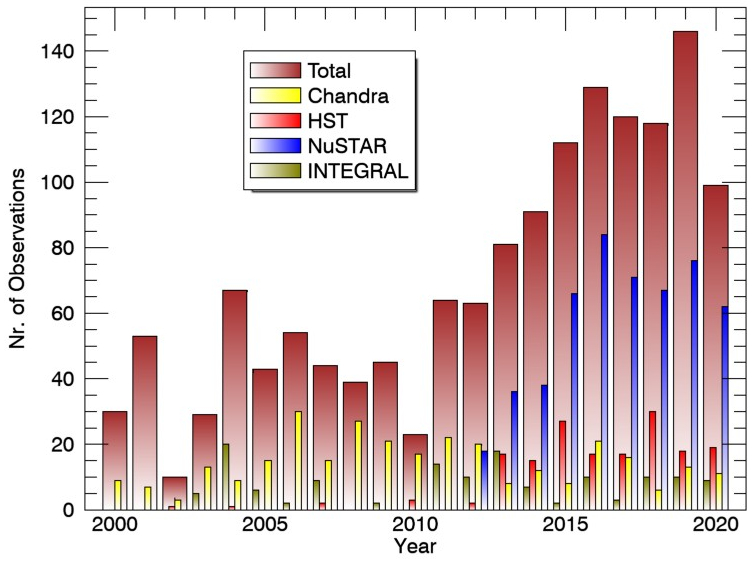}
		\caption{\label{Fig:joint}
 Observations performed coordinated or simultaneous with other facilities}	 
	\end{center}
    \end{minipage}
\end{figure}

\begin{figure}
  \begin{minipage}[t]{5.5cm}
    	\begin{center}
		\includegraphics[angle=0,scale=0.25]{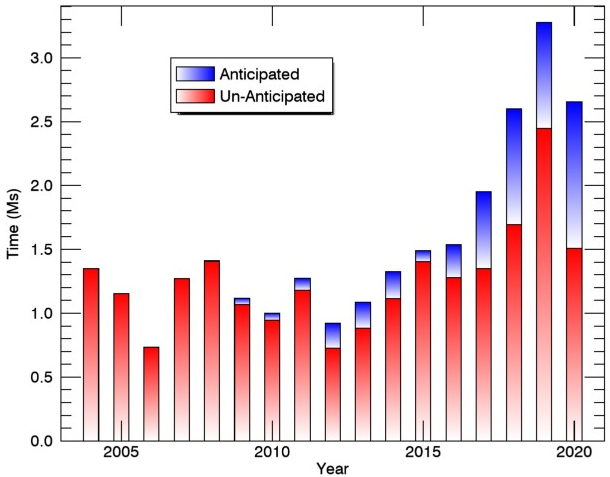}
		\caption{\label{Fig:too2}
 Observing time performed for Target of Opportunity observations
}
	\end{center}
    \end{minipage}
  \hfill
  \begin{minipage}[t]{5.5cm}
    	\begin{center}
	\includegraphics[angle=0,scale=0.19]{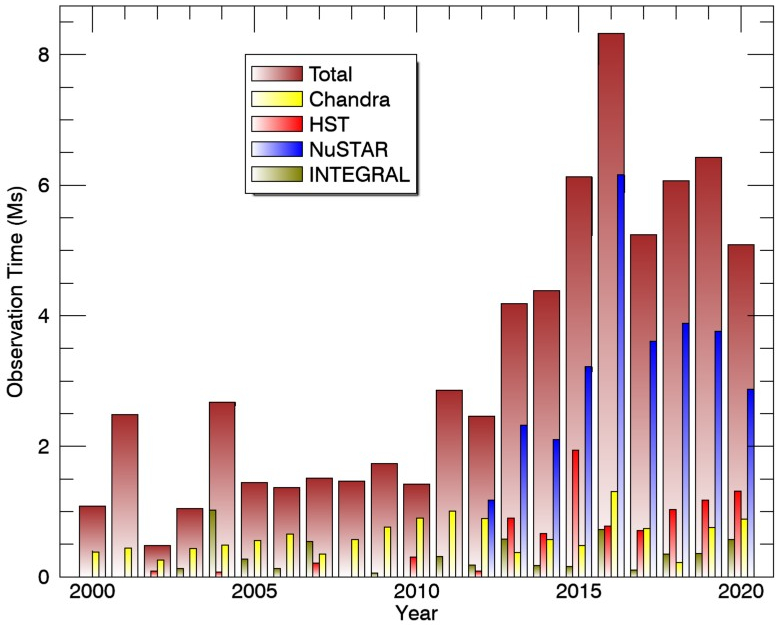}
		\caption{\label{Fig:joint2}
 Observing time performed coordinated or simultaneous with other facilities}	 
	\end{center}
    \end{minipage}
\end{figure}

About 15\% of submitted proposals request 
anticipated target of opportunity (TOO) observations.
In addition there are about 45 requests for unanticipated 
TOOs and/or Director's discretionary time observations 
each year, which in general are forwarded
to an OTAC (Observing Time Allocation Committee) 
chairperson for a recommendation.
XMM-Newton has not formally limited the time available for TOOs.
However, the high amount of fixed-time observations 
and observations performed simultaneously with other 
missions limit the time which may be allocated to TOO 
observations. In the very late 2010s and very early 2020s  
XMM-Newton typically performed 80 TOO observations per year. 
This rate is more than double the rate of TOO observations in the 
early days of the mission (e.g., 2005); see Fig.~\ref{Fig:too} and  Fig.~\ref{Fig:too2}.

Multi-wavelength and multi-messenger observations are 
a powerful tool to foster transformative science.
With its suite of instruments, XMM-Newton already provides multi-wavelength coverage from the optical to X-ray regimes. 
XMM-Newton SOC staff investigated ways to further extend this approach to "multi"-wavelength
observations. These can be performed via joint programs.
In 2021, XMM-Newton has joint programs with nine facilities.
The joint programs allow the time allocation committees of each facility
to allocate time on the other mission in connection with the allocation
of time on the own facility.
Therefore, with one proposal, in response to an XMM-Newton AO, observing
time on up to 10 facilities can be requested.
Table~\ref{tab:joint} shows the joint programs and the time exchanged per year.
The XMM-Newton joint programs allow the spectral energy
distribution of a source to be covered from the radio, optical, 
UV and X-ray, $\Gamma$-ray all the way to the TeV range.
About 30\% of the performed high-priority observations 
(i.e. observations whose execution is guaranteed) 
are from a joint program, most of them simultaneous with
one or more facilities; see Fig.~\ref{Fig:joint} and  Fig.~\ref{Fig:joint2}.

\begin{table}   
\begin{center}
\begin{tabular}{|l|c|} \hline
Facility  &  Exchanged time \\ \hline \hline
NRAO      &   2 $\times$ 150 ks    \\
VLT(I)    &   2 $\times$ 290 ks \\
HST       &   2 $\times$ 150 \\
Chandra   &   2  $\times$ 1Ms  \\
Swift     &   300  ks \\
NuSTAR    &   2  $\times$ 1.5 Ms \\
INTEGRAL  &   2  $\times$ 300 ks \\
MAGIC     &   2  $\times$ 150 ks \\
H.E.S.S.  &   2  $\times$ 150 ks  \\ \hline
\end{tabular}
\end{center}
	\caption{{Joint Programmes.}}
	\label{tab:joint}
\end{table}

\begin{figure}
  \begin{minipage}[t]{5.5cm}
    	\begin{center}
		\includegraphics[angle=0,scale=0.15]{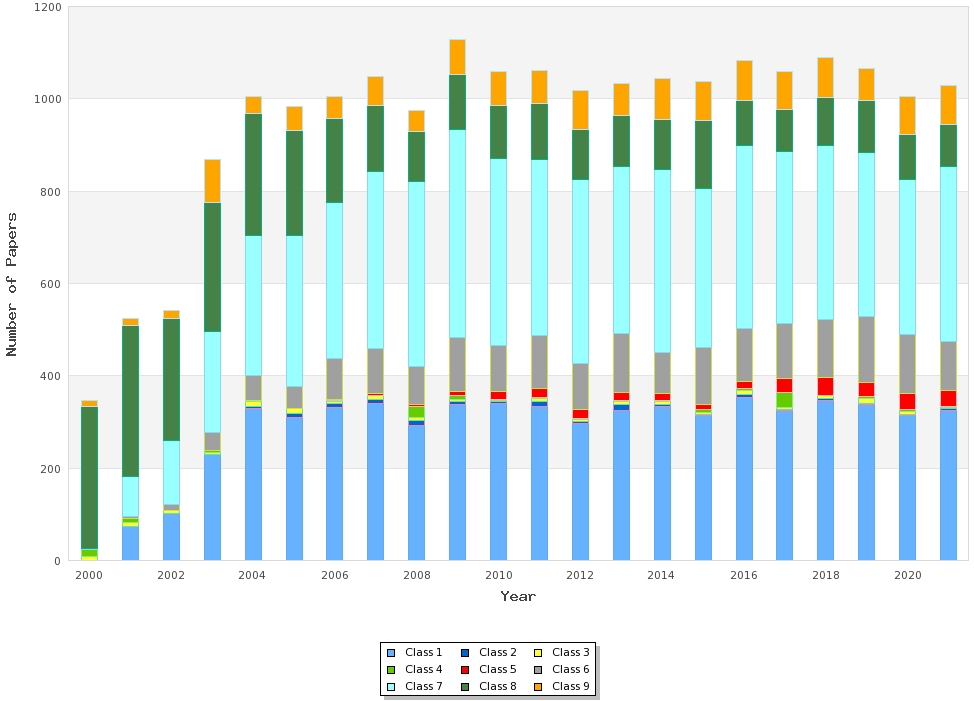}
		\caption{\label{Fig:pub}
Classification of all refereed articles listed in ADS which contain
"XMM" in a full text-search:
(1) article makes use of XMM-Newton data or pipeline products;
(2) catalogue based on XMM-Newton observations;
(3) article makes quantitative predictions for XMM-Newton observations;
(4) article describes XMM-Newton, its instruments, scientific impact, etc.
(5) article makes use of the primary catalogues
(6) article makes use of published XMM-Newton results
(7) article refers to papers presenting XMM-Newton results
(8) article refers to "XMM-Newton" in general
(9) article uses expression derived from XMM-Newton, e.g., names of objects
}
	\end{center}
    \end{minipage}
  \hfill
  \begin{minipage}[t]{5.5cm}
    	\begin{center}
	\includegraphics[angle=0,scale=0.15]{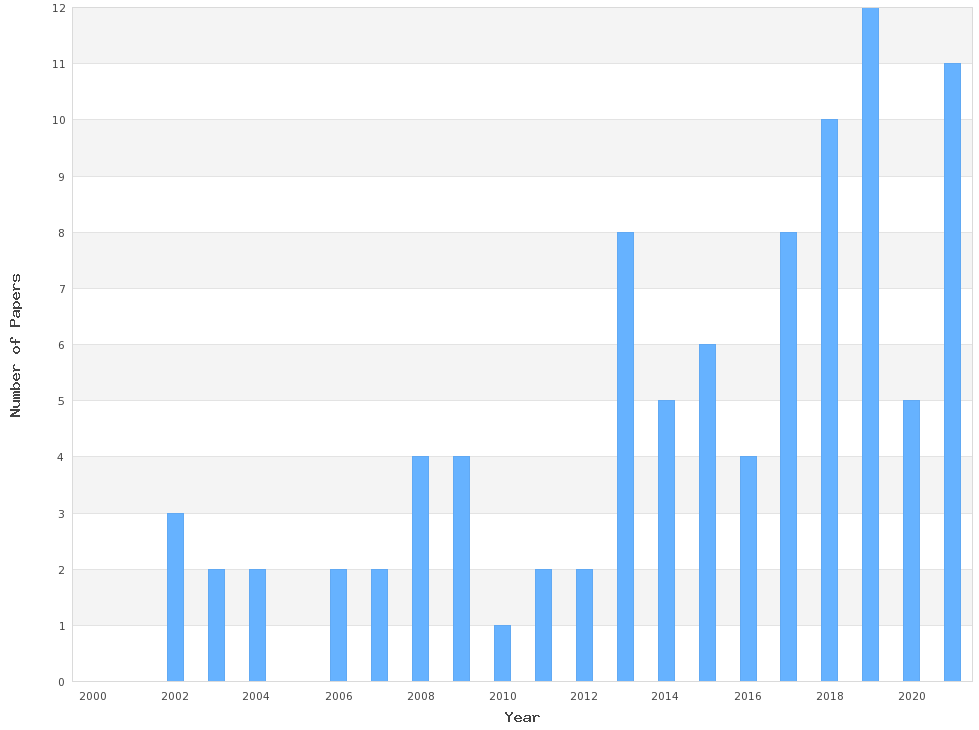}
	\caption{\label{Fig:pubhigh}
          Articles in {\it Nature} and {\it Science} making use of XMM-Newton data 
are shown per year as an indicator of the amount of 
transformative science results}	 
	\end{center}
    \end{minipage}
\end{figure}

Requests for observing time longer than 300 ks are submitted 
as large programs. About 40\% of the high-priority observing 
time (execution guaranteed) is given to large programs,
such that they have the same over-subscription as the normal programs.
Examples of transforming science resulting from larger programs are
given in \cite{Fabian2009}, \cite{Alston2020}, \cite{Nicastro2018}
To accomplish programmes requiring more than 2 Ms, e.g., \cite{Pierre2017}, the
call for Multi-Year-Heritage (MYH) program was introduced in 2017.
Here, up to 6 Ms of observing time are available for distribution, 
to be executed over a period of three AOs (3 years).
The call for MYH programs in 2017 had an over-subscription of 
a factor 10. The call in 2020 suffered the pandemic constraints  
showing a lower over-subscription.

The distribution of elapsed time between performed
observations and publication peaks at 2 years \cite{Ness2014} .
Most of the times, TOO observations lead to rapid publications 
with about 1 year elapsed time between observation and 
publication \cite{Ness2014}. 
The predefined observing modes in combination with the provided 
support, e.g., pipeline products, calibration, archive, or catalogues,
make XMM-Newton data highly comparable and therefore ideally suited for
studies based on archival data. In fact 90\% of the 
observing time has been used in at least one publication \cite{Ness2014}.

XMM-Newton data also play an important role in education
and in the development of new generations of researchers around the world.  
At the time of writing (November 2021), 406 Ph.D. theses had used XMM-Newton
data or included results of research related to the development of the instruments.
XMM-Newton is frequently used by young scientists who are starting
their career in astrophysics. 
Since the 5th call for observing time proposals, in 2005, astronomers 
sending proposals the first time submit about 20\% of the proposals
of each call. The success rate of these first-time proposers is slightly
below the average success rate of all proposers. Remarkably, 
in more than one third of the calls since 2005, the rate of requested
to allocated observing time secured by first-time proposers was similar
or even larger than the average rate of all proposers.

About 380 articles are 
published in refereed journals each year making
use of XMM-Newton data, describing the instruments, or
using pipeline products or the catalogue.
Figure \ref{Fig:pub} gives an analysis of all refereed articles 
listed in ADS, which mention XMM-Newton, demonstrating not 
only the usage of XMM-Newton data but also its impact via 
citations. Articles containing results based on XMM-Newton 
observations are about three times more cited than all 
astronomical papers. 
Fig.~\ref{Fig:pubhigh}  shows the number of articles making use of XMM-Newton
data published in {\it Nature} or {\it Science} journal  demonstrating the
high amount of transforming science which typically is published
in these two journals.


\section{\textit{Authors Contribution}}

Norbert Schartel and Maria Santos-Lle{\'o} contributed  the "Introduction", 
"Scientific Data and Analysis", and "Scientific Strategy and Impact" 
sections, Rosario Gonz{\'a}lez-Riestra contributed  "The Reflection Grating 
Spectrometers (RGSs)" section, Peter Kretschmar contributed  the "Organization of 
the XMM-Newton Ground Segment" section, Marcus Kirsch contributed  "The 
Spacecraft" section, Pedro Rodr{\'i}guez contributed  the "Observing with 
XMM-Newton" section, Simon Rosen contributed the "Optical Monitor (OM)" section,  
Michael Smith and Martin Stuhlinger contributed  the "European Photon Imaging 
Camera (EPIC)" section and Eva Verdugo-Rodrigo contributed  the "X-ray Mirrors" 
section. Norbert Schartel prepared the chapter outline compiled and 
homogenized the different contributions.

\section{\textit{Acknowledgments}}
 The authors deeply thank Arvind Parmar for many useful suggestions and
 Lucia Ballo for help with the statistical numbers for joint 
programs. The authors also acknowledge the outstanding contribution of 
the Survey Science Centre to the mission success and the dedication and 
excellence of the rest of the team members of the XMM-Newton 
Mission Operations and Science Operations centres, where the authors of 
this paper feel as an honour to serve.

\end{document}